\documentclass{elsarticle}
\usepackage{graphics}
\usepackage{epsfig}
\usepackage{color}
\usepackage{lineno,amssymb}
\usepackage{multirow}
\usepackage{wasysym}        
\usepackage{hyperref}

\begin{document}


\begin{frontmatter}



\title{The Fluorescence Detector of the Pierre Auger Observatory}



\author{\noindent{\bf The Pierre Auger Collaboration} \\
J.~Abraham$^{8}$, 
P.~Abreu$^{71}$, 
M.~Aglietta$^{54}$, 
C.~Aguirre$^{12}$, 
E.J.~Ahn$^{87}$, 
D.~Allard$^{31}$, 
I.~Allekotte$^{1}$, 
J.~Allen$^{90}$, 
P.~Allison$^{92}$, 
J.~Alvarez-Mu\~{n}iz$^{78}$
% ...
\\
\noindent $^{1}$ Centro At\'{o}mico Bariloche and Instituto Balseiro (CNEA-UNCuyo-CONICET), San Carlos de Bariloche, Argentina \\
% ...
$^{8}$ Universidad Tecnol\'{o}gica Nacional, Facultad Regional Mendoza, 
(UTN-FRM), Mendoza, Argentina \\
% ...
$^{12}$ Universidad Catolica de Bolivia, La Paz, Bolivia \\
% ...
$^{31}$ Laboratoire AstroParticule et Cosmologie (APC), 
Universit\'{e} Paris 7, CNRS-IN2P3, Paris, France \\
% ...
$^{54}$ Istituto di Fisica dello Spazio Interplanetario (INAF),
 Universit\`{a} di Torino and Sezione INFN, Torino, Italy \\
% ...
$^{71}$ LIP and Instituto Superior T\'{e}cnico, Lisboa, Portugal \\
% ...
$^{78}$ Universidad de Santiago de Compostela, Spain \\
% ...
$^{87}$ Fermilab, Batavia, IL, USA \\
% ...
$^{90}$ New York University, New York, NY, USA \\
% ...
$^{92}$ Ohio State University, Columbus, OH, USA \\
% ...
\noindent (\ddag) Deceased \\
(a) at Konan University, Kobe, Japan \\
(b) On leave of absence at the Instituto Nacional de Astrofisica, Optica y Electronica \\
(c) at Caltech, Pasadena, USA \\
}



\begin{abstract}
The Pierre Auger Observatory is a hybrid detector for ultra-high
energy cosmic rays. 
It combines a surface array to
measure secondary particles at ground level together with a
fluorescence detector to measure the development of air
showers in the atmosphere above the array.  The fluorescence
detector comprises 24 large telescopes specialized for
measuring the nitrogen fluorescence caused by charged
particles of cosmic ray air showers. 
In this paper we describe the components
of the fluorescence detector including its optical system, the
design of the camera, the electronics, and the systems for
relative and absolute calibration. We also discuss the operation
and the monitoring of the detector. Finally, we evaluate the
detector performance and precision of shower reconstructions.
\end{abstract}

\end{frontmatter}
\newpage


\section{Introduction}

\begin{figure}[ht]
\begin{center}
\includegraphics [width=0.6\textwidth]{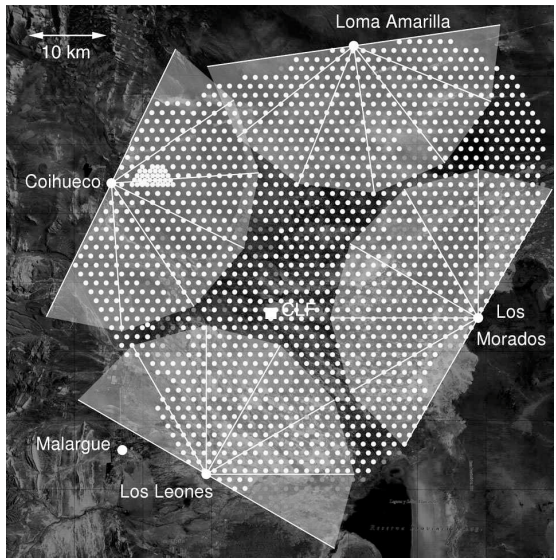}
\caption{Status of the Pierre Auger Observatory as of March 2009.
Gray dots show the positions of surface detector 
stations, lighter gray shades indicate deployed detectors, while
dark gray defines empty positions. Light gray segments indicate
the fields of view of 24 fluorescence telescopes which are
located in four buildings on the perimeter of the surface array.
Also shown is a partially completed infill array near the Coihueco 
station and the position of the Central Laser Facility (CLF, 
indicated by a white square). The description of the CLF 
and also the description of all other atmospheric monitoring 
instruments of the Pierre Auger Observatory is available 
in \cite{sybenzvi}.} 
\label{array}
\end{center}
\end{figure}

The hybrid detector of the Pierre Auger Observatory \cite{general} 
consists of
1600 surface stations -- water Cherenkov tanks and their
associated electronics -- and 24 air fluorescence telescopes. 
The Observatory is located outside the city of Malarg\"{u}e,
Argentina (69$^{\circ}$ W, 35$^{\circ}$ S, 1400 m a.s.l.) and the
detector layout is shown in Fig. \ref{array}. Details of the
construction, deployment and maintenance of the array of 
surface detectors
are described elsewhere \cite{SD}. In this paper we will
concentrate on details of the fluorescence detector and its performance.

The detection of ultra-high energy ($\gtrsim 10^{18}$~eV)
cosmic rays using nitrogen fluorescence emission
induced by extensive air showers is a well
established technique, used previously by the Fly's Eye \cite{flyseye} 
and HiRes \cite{hires} experiments. It is used also for the Telescope Array \cite{telarray} 
project that is currently under construction, and it has been proposed 
for the satellite-based EUSO and OWL projects. 

Charged particles generated during the development 
of extensive air showers excite atmospheric nitrogen molecules, 
and these molecules then emit fluorescence light in the $\sim 300-430$~nm
range. 
The number of emitted fluorescence photons is proportional to the
energy deposited in the atmosphere due to electromagnetic energy
losses by the charged particles.  By measuring the rate of
fluorescence emission as a function of atmospheric slant depth $X$, an
air fluorescence detector measures the {\it longitudinal development
profile} $\frac{dE}{dX}(X)$ of the air shower.  The integral of this
profile gives the total energy dissipated electromagnetically, which
is approximately 90\% of the total energy of the primary cosmic ray.


For any waveband, the fluorescence yield is defined as the number of
photons emitted in that band per unit of energy loss by charged particles.
The absolute fluorescence yield in air at 293~K and 1013~hPa from the
337~nm fluorescence band is $5.05 \pm  0.71$ photons/MeV of energy deposited, 
as measured in \cite{nagano}. The wavelength dependence of the
yield has been described e.g. in \cite{airfly}. 
Since a typical cosmic ray shower spans over 10 km in altitude, it is important
to stress that due to collisional quenching effects the fluorescence yield
is also dependent on pressure, temperature and humidity of the air.

\begin{figure}[t]
\begin{center}
\includegraphics [width=0.9\textwidth]{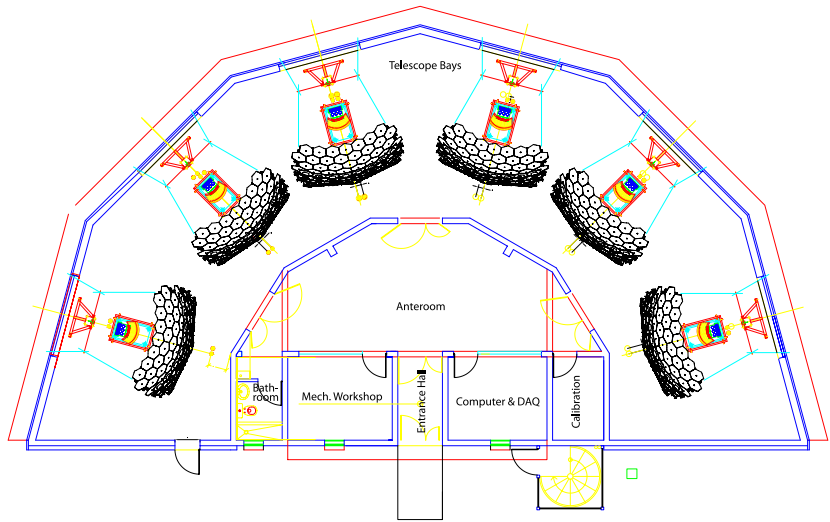} 
\caption{Schematic layout of the building with six fluorescence telescopes.} 
\label{fdeye}
\end{center}
\end{figure}

The fluorescence detector (FD) comprises four observation sites --- 
Los Leones, Los Morados, Loma
Amarilla, and Coihueco --- located atop small elevations on the
perimeter of the SD 
array. Six independent telescopes, each with field of view of
$30^{\circ} \times 30^{\circ}$ 
in azimuth and elevation, are located in each FD site. 
 The telescopes face towards the interior of the array so that the
 combination of 
the six telescopes provides $180^{\circ}$ coverage in azimuth. Figure
\ref{fdeye} shows the arrangement of the telescopes inside an observation site.

Figure 3 depicts an individual FD telescope. The telescope is housed in a clean
climate-controlled building.  Nitrogen fluorescence light enters through a large 
UV-passing filter window and a Schmidt optics corrector ring. The light is focused 
by a 10 square meter mirror onto a camera of 440 pixels with photomultiplier light 
sensors.  Light pulses in the pixels are digitized every 100 nanoseconds, and 
a hierarchy of trigger levels culminates in the detection and recording of
cosmic ray air showers.


\begin{figure}[ht]
\begin{center}
\includegraphics [width=0.5\textwidth]{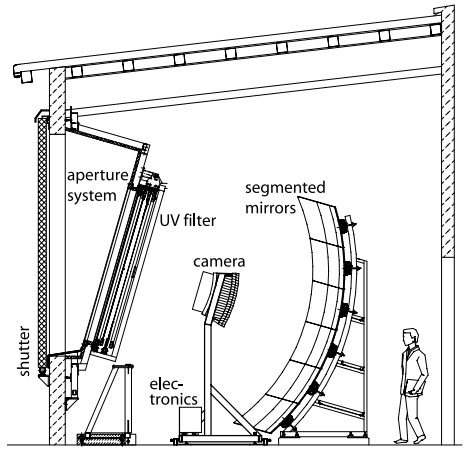} 
\caption{Schematic view of a fluorescence telescope of the Pierre
Auger Observatory.} 
\label{fig1}
\end{center}
\end{figure} 

This paper is organized as follows. In section
\ref{opticalsystem} we describe the components of the optical
system of an individual telescope, and in section
\ref{telescopecamera} we focus on the telescope camera. The
electronics of a fluorescence telescope and the data acquisition system (DAQ) 
of an FD station are described in section \ref{electronicsystem}. The
details of the calibration hardware and methods are given in
section \ref{calibration}, and the performance, operation and
monitoring of the fluorescence detector are explained in section
\ref{performance}. Finally, in section \ref{reconstruction} we
concentrate on the basics of shower reconstruction using the
measured fluorescence signal, and in section \ref{conclusions} we
summarize.

\section{Optical system}\label{opticalsystem}

\begin{figure}[h!]
\begin{center}
\includegraphics[width=0.6\textwidth]{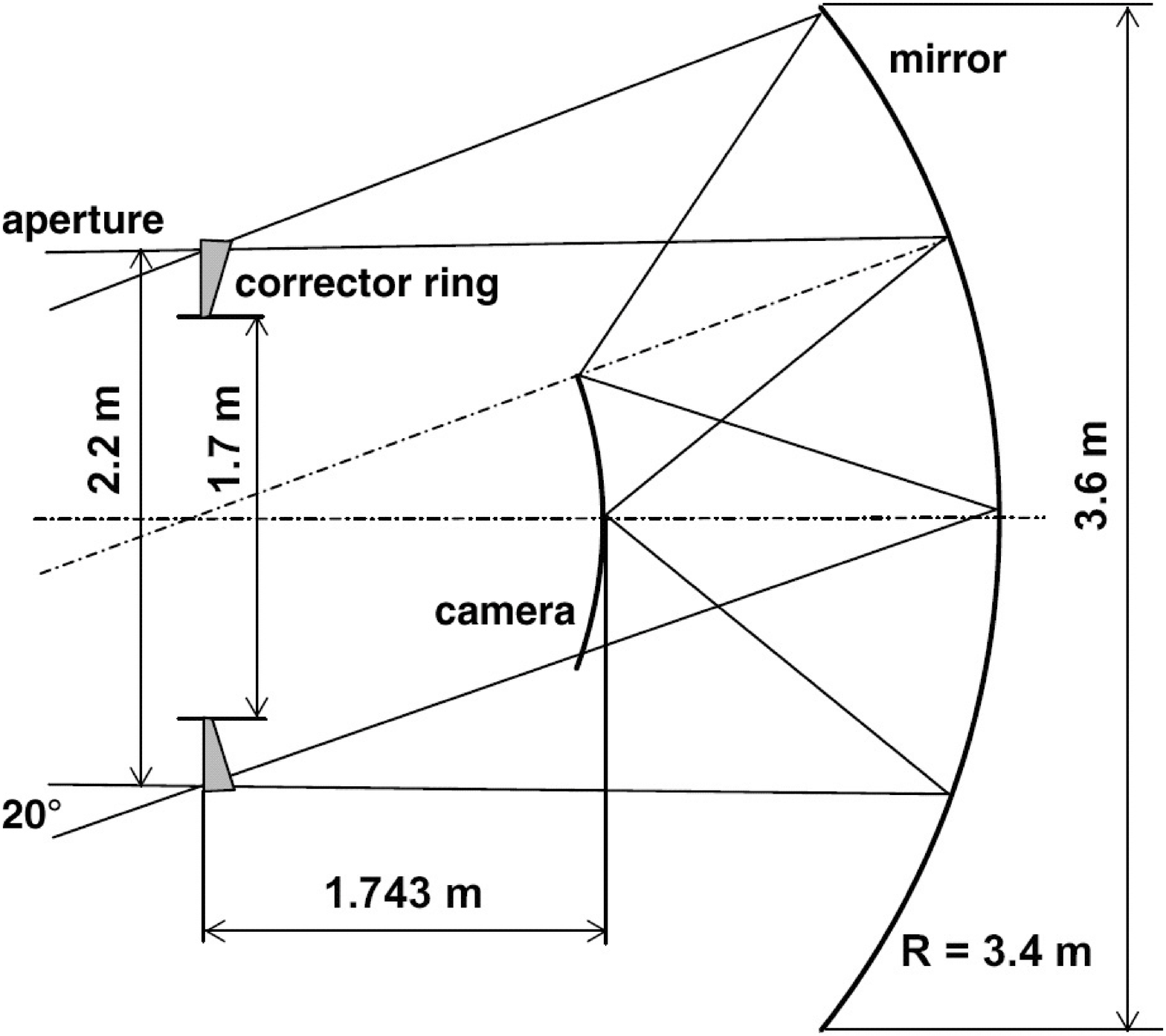}
\caption{Geometrical parameters of the FD telescopes.}
\label{fig4}
\end{center}
\end{figure}

The basic elements of the optical system in each
FD telescope are  a filter at the entrance
window, a circular aperture, a corrector 
ring, a mirror and a camera with photomultipliers. The geometrical
layout of the components is depicted in Fig.~\ref{fig4}. 

The window is an
optical filter made of Schott MUG-6 glass \cite{Schott}.  This absorbs
visible light while transmitting UV photons up to 410 nm
wavelength, which includes almost all of the nitrogen
fluorescence spectrum.  Without the filter window, the
fluorescence signals would be lost in the noise of visible
photons.

\begin{figure}[h!]
\begin{center}
\includegraphics[width=0.8\textwidth]{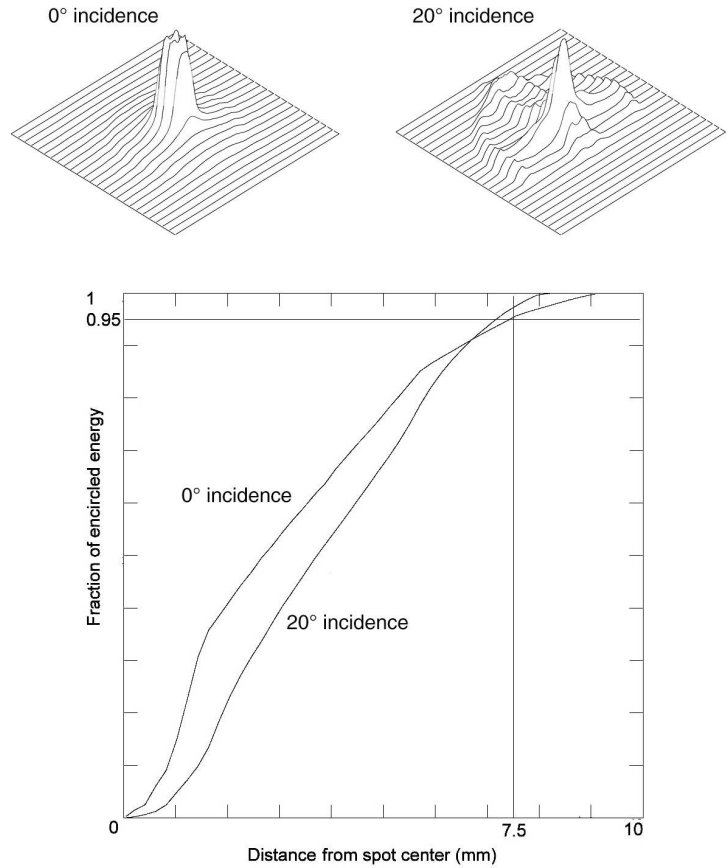}
\caption{ \emph{Top:} The simulated 3D distribution of the light
intensity for spots on the optical axis (left) and close to the
camera corner (right; distance from the optical axis =
$20^{\circ}$). The size of the imaged area is $20 \times 20$ mm.
\emph{Bottom:} Fraction of encircled energy as a function of spot
diameter for the spot on the optical axis (upper curve; angular
distance from the optical axis = $0^{\circ}$; this curve
corresponds to the distribution shown top left) and for the spot
close to the camera corner (lower curve; angular distance from
the optical axis = $20^{\circ}$; this curve corresponds to the
distribution shown top right).
}
\label{fig5}
\end{center}
\end{figure}

The aperture, the corrector ring, the mirror, and the PMT camera
constitute a modified Schmidt camera design that partially corrects
spherical aberration and eliminates coma aberration.
The size of the aperture is optimized to keep the spot size\footnote{The 
image of the point source at
infinity on the focal surface of the optical system is commonly
called the ``spot'' in optics, but it may be better known as a
``point spread function''. The size of the spot characterizes the 
quality of the optical system.} due
to spherical aberration within a diameter of 15 mm, i.e.\ 90\% 
of the light from a distant point source located anywhere within 
the $30^{\circ} \times 30^{\circ}$ FOV of a camera falls into a circle 
of this diameter. This corresponds to an angular spread of $0.5^{\circ}$. 
In comparison, the FOV of a single camera pixel is $1.5^{\circ}$. The light 
distribution within the spot is described by the point spread 
function (PSF) shown in Fig.\ \ref{fig5}.

\begin{figure}[h!]
\begin{center}
\includegraphics[width=0.8\textwidth]{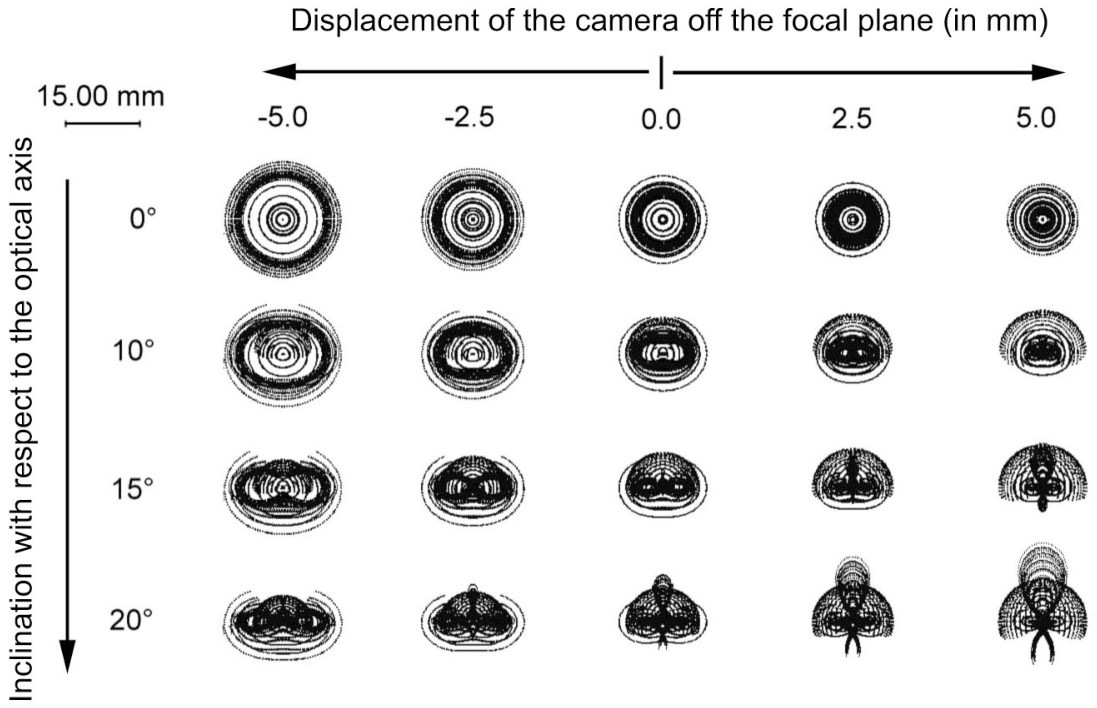}
\caption{Spot diagrams of the telescope.}
\label{fig7}
\end{center}
\end{figure}

The schematic view of the spot size diagrams over the whole FOV
is shown in Fig.\ \ref{fig7}, where the rows correspond to
viewing angles $0^{\circ}$, 10$^{\circ}$, $15^{\circ}$ and
$20^{\circ}$. The columns corresponds to different displacements of the 
camera off the focal plane, by
changing the camera-mirror distance from $-5$~mm to $+5$~mm with
respect to the nominal separation. The central position is
located at a distance of  1\,657~mm from the primary mirror. The
asymmetric shape of some  spots is due to vignetting and camera shadow. This
picture also shows the sensitivity of the telescope PSF to the
precision of the adjustment of the distance between mirror and
camera. 

\subsection{Segmented mirror}

Due to the large area of the primary mirror ($\sim13$~m$^2$), the mirror is segmented to reduce
the cost and weight of the optical system.  Two segmentation configurations
are used: first, a tessellation of 36 rectangular anodized aluminum mirrors of three different 
sizes; and second, a structure of 60 hexagonal glass mirrors (of four shapes and sizes) with 
vacuum-deposited reflective coatings.  In both cases, all mirror segments have a spherical
inner radius of 3400~mm, allowing possible deviations up to 3420~mm.



\begin{figure}[h!]
\begin{center}
\includegraphics[width=0.8\textwidth]{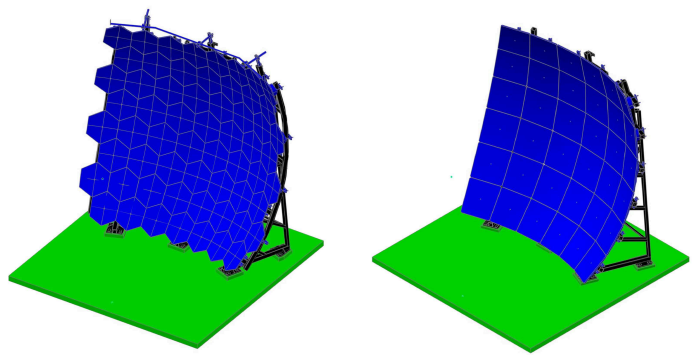}
\caption{Two different segmentation configurations of the telescope mirror: 
{\emph left:} 36 rectangular anodized 
aluminum mirror segments; {\emph right:} 60 hexagonal glass mirror segments.}
\label{tessellations}
\end{center}
\end{figure}

The 12  telescopes at the Los Leones and Los Morados sites use aluminum
mirrors.  The mirror elements were produced from 20~mm-thick cast aluminum blocks, and were
milled to the required spherical inner radius using a diamond milling technique.  After the
initial milling, the reflective surface was created by gluing a 2~mm sheet of AlMgSiO$_5$
alloy onto the inner surface of each element.  The sheets were attached to the aluminum block
at elevated temperature and pressure to achieve high stability.  The material used is 
well-suited for high-precision diamond milling, and allows the final mirror surface to 
achieve a roughness below 10~nm.  Finally, a 90~nm thick aluminum-oxide layer was applied to
the surface by chemical anodization to provide additional protection.

The remaining 12 telescopes in the buildings at 
Coihueco and Loma Amarilla use mirrors with glass segments. The thickness 
of the glass is 12~mm, and it is composed of SIMAX, a borosilicate glass of the
PYREX type.  The choice of ultra-thin, lightweight glass was motivated
by the need to maintain the optical stability of the segments at different temperatures
and for different inclinations of the individual segments.  SIMAX is suitable for machine
grinding and polishing, and has good mechanical and thermal stability.


The reflective layer on the glass mirror segments is composed of Al layer, 
with thickness of 90~nm, covered by one SiO$_{2}$ layer with thickness of
110~nm.  The aluminum is used 
for its high reflectivity, and the silica for its high mechanical resistance. 
The reflectivity has been measured at several points of each segment, and the average
reflectivity at $\lambda=370$~nm exceeds 90\%. The surface roughness has been measured 
mechanically and optically, and has an RMS of less than 10~nm.

In the glass mirrors, the mirror segments are grouped by their curvature radii to minimize
the radius deviations in each telescope.  
The curvature and the shape of the reflective surface 
of the segments are measured and controlled using standard Ronchi and/or 
Hartmann tests \cite{malacara}. 

The alignment of the mirror segments is accomplished by directing a
laser  onto  
each segment. The laser is mounted on the center of curvature of the full
mirror.  The laser beam approximates a point source, and each segment
is adjusted 
such that the beam is reflected to a common point (the center of the laser aperture).  This
procedure is used to adjust all the mirror segments and match the full mirror to the required
spherical shape, with the center of curvature aligned with the optical axis of the telescope.




\subsection{Corrector ring}


A novel solution of the optical system with a ``corrector ring'' was designed to
keep the advantage of a large aperture of the Schmidt system, 
and simultaneously simplify the production of the 
element,  minimize its weight and cost, while maintaining 
the spot size within the limits of aforementioned design specification
\cite{lenses1}.  The aperture area of the telescope with the corrector ring
is almost doubled with respect to the optical 
system without any correcting element.
The analysis of real shower data \cite{lenses2} has compared the performances of FD 
optical systems with and without corrector ring\footnote{Test measurements without corrector ring
were realized on several FD telescopes 
prior to corrector ring installation.}, and has verified that the
corrector ring enhances the FD aperture by a factor of $\sim2$.

\begin{figure}[ht]
\begin{center}
\includegraphics[width=0.8\textwidth]{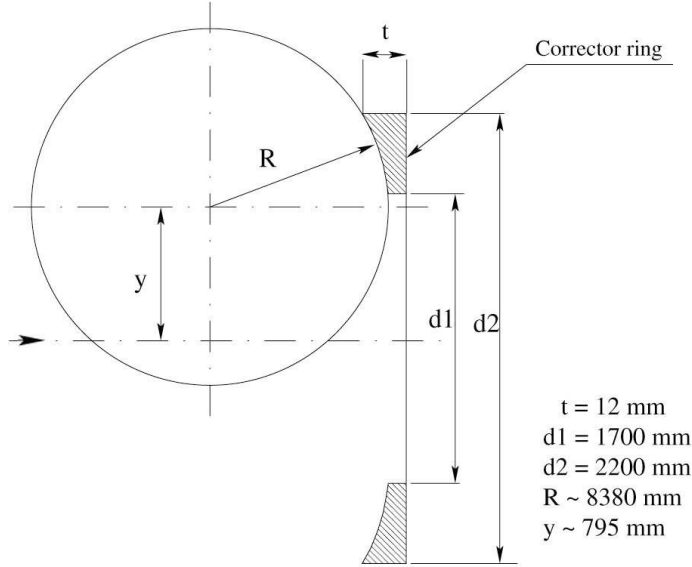} 
\end{center}
\caption{Corrector ring profile using a spherical approximation, the optical axis
of the telescope is identified by the dash-dotted line with 
an arrow on the left and $d2$ is the diameter of the aperture.}
\label{fig3}
\end{figure}

The corrector ring is the circumferential part of the
corrector plate of a classical Schmidt camera with one planar
side and the other with an aspheric shape corresponding to a 
$6^{\mbox{\small th}}$-order polynomial curve. 
Such a surface is difficult to manufacture and therefore some
optimizations were adopted to simplify the lens production.
Eventually, a spherical approximation specially designed for the
fluorescence detector was chosen to fulfill both price and
performance requirements (see Fig.\ \ref{fig3}). 
The simplified corrector ring is located at the aperture and
covers the annulus between radii of 0.85 m and 1.1 m. 


\begin{figure}[tb]
\begin{center}
\includegraphics[width=9cm]{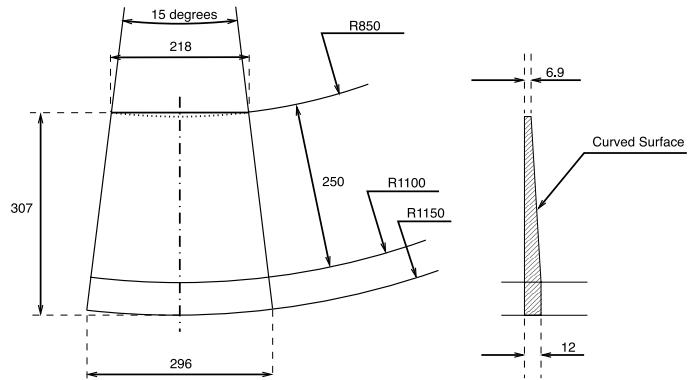} 
\caption{Technical drawing of one corrector ring segment. The dimensions are given in millimeters.}
\label{segment}
\end{center}
\end{figure}

Since the rings have an external diameter of 2.2~m,
their manufacture and transportation to the site in a single
piece would have been very difficult. Therefore, each lens was 
divided into 24 segments. The size and profile 
of one segment is shown in Fig.\ \ref{segment}. 
The production of the segments has been performed by
Schwantz Ltd \cite{Schwantz} after assembling a machine with a
circular base to hold the segments, and a disk with diamond
abrasive cylinders for the grinding of the glass (BK7 glass from
Schott \cite{Schott}) to the desired profile.
After production of each ring, its segments were tested for
proper shape. To scan the ring profiles, a laser beam was pointed towards the
curved and flat surfaces and the positions of the reflected light from each
surface were measured \cite{lenses1}. 

\subsection{Simulation of the optical system}

To evaluate the overall optical performance of the detector and
to confirm the theoretical expectations, a dedicated complete
simulation of the optical system
was developed using Geant4 \cite{sim1, sim2}, a Monte Carlo
toolkit for the simulation of radiation and light propagation.  
The tracking of optical photons includes refraction and
reflection at medium boundaries, Rayleigh scattering and bulk
absorption. The optical properties of a medium, such as
refractive index, absorption length, and reflectivity
coefficients, can be expressed as functions of the wavelength.
The application includes a detailed description of the different
detector elements -- UV filter, corrector lens, mirror and
camera. The optical properties of all materials, such as the
absorption length and the refractive index, were implemented as a
function of wavelength.

The simulation confirms that all design specifications of the FD
optical system are met, i.e. even in the corners of the camera 90\% 
of the light from a distant point source is concentrated within a diameter of 15 mm.


\begin{figure}[ht]
\begin{center}
\includegraphics[width=0.8\textwidth]{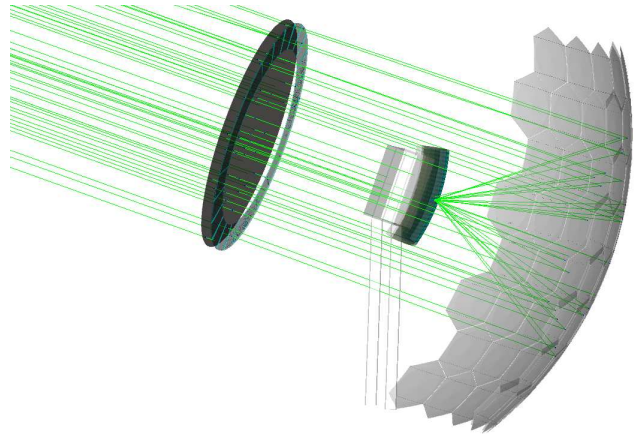}
\caption{Ray tracing simulation of the optical system of the telescope made using Geant4.}
\label{geant4}
\end{center}
\end{figure}



\section{Telescope camera}\label{telescopecamera}


The camera \cite{camera} is the sensitive element of a telescope.
It is composed of a matrix of 440 pixels located on the focal
surface of the telescope. The camera pixels are arranged in a
matrix of 22 rows by 20 columns (Fig.~\ref{figcamera1}c). 
The corresponding field of view is of 30$^\circ$ in azimuth 
(full acceptance of one row) and 28.1$^\circ$ 
in elevation (full acceptance of one column). 

A cosmic ray shower is imaged on the camera as a line of
activated pixels having a track-like geometrical pattern and also
a clear time sequence. Each pixel is realized by a
photomultiplier with a light collector.
      
\subsection {Geometry}

\begin{figure}[htb]
\begin{center}
\vspace*{-1.5cm}
\begin{center}\epsfig{file=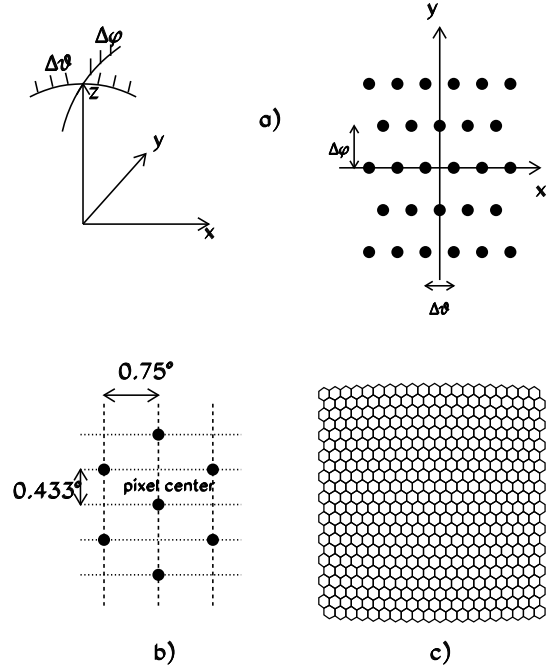,height=12.cm}
\end{center}
\vspace*{-2.cm}
\caption{Geometrical construction of the FD camera; a) pixel centers are placed 
on the spherical surface in steps of $\Delta \theta = 1.5^{\circ}$ 
and $\Delta \phi = 0.866^{\circ}$, b)
positioning of the pixel vertices around the pixel center, c) the camera with
 440 pixels arranged in a 22x20 matrix. }
\label{figcamera1}
\end{center}
\end{figure}

The pixel array lies on the focal surface of the optical system,
which is a sphere of 1.743 m radius.  
The pixels are hexagonal with a side to side distance of 45.6~mm, corresponding to an angular size of 1.5$^\circ$. The pixel
centers are placed on the spherical surface following the
procedure outlined in Fig.~\ref{figcamera1}, where we define
$\Delta \theta = 1.5^{\circ}$ and $\Delta \phi = 0.866^{\circ}$.
The first center is placed at +$\Delta \theta/2$ with respect to
the telescope axis, which is taken as the {\it z} axis in
Fig.~\ref{figcamera1}. The other pixel centers are obtained with
increasing (or decreasing) $\Delta \theta$ steps. In this manner,
a row of twenty pixels (corresponding to 30$^{\circ}$ in azimuth)
is built.

The following row of pixels is obtained by a rotation of $\Delta
\phi$ around the {\it x} axis. To produce the correct staggering
between rows, the pixel centers are moved by $\Delta \theta/2$
with respect to their positions in the previous row.  

Once the pixel centers have been defined, each pixel hexagon is
determined by positioning six vertices around the center. The
angular positions of the vertices are obtained by moving in steps
of $\Delta \theta/2$ and $\Delta \phi/3$ with respect to the
pixel center, as depicted in Fig.~\ref{figcamera1}b. Equal steps
in angle produce different linear dimensions depending on the
pixel position on the spherical surface. Therefore, the pixels
are not exactly regular hexagons, but their size varies over the
focal surface by 3.5\% at most. Differences in the side length
are smaller than 1 mm, and are taken into account in the design
of the light collectors (see section \ref{2.3}) and in the
analysis.

 \subsection {Mechanics}
\begin{figure}[htb]
\begin{center}
\epsfig{file=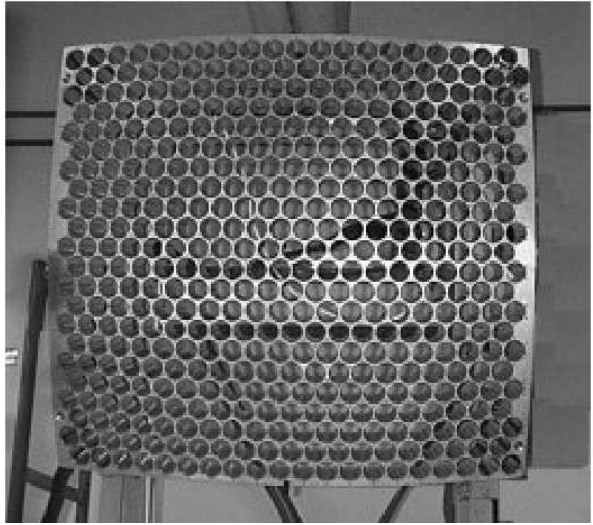,height=7.cm}
\caption{Picture of the camera body.}
\label{figcamera2}
\end{center}
\end{figure}

The camera body was produced from a single aluminum block by a
programmable milling machine. It consists of a plate of
60 mm uniform thickness and approximately rectangular shape (930
mm horizontal $\times$ 860 mm vertical), with spherical outer and
inner surfaces. The outer radius of curvature is 1701~mm, while
the inner radius is 1641~mm. Photomultiplier tubes are positioned
inside 40 mm diameter holes drilled through the plate on the
locations of the pixel centers. Small holes in the camera body at
the pixel vertices are used to secure the light collectors in
place.  A picture of the camera body is shown in
Fig.~\ref{figcamera2}.

Within a finished telescope, the camera body is held in place by
a simple and robust two-leg steel support made of 5 cm wide
C-shaped steel beams. The obscuration of the mirror FOV due to
the camera support is less than one-tenth of that of the camera.
Power and signal cables run inside the C-shaped legs of the
support without producing additional obscuration.

\subsubsection{Mechanical precision and alignment}

To align an FD telescope, the pixel surface must be placed at the
correct longitudinal distance from the center of curvature of the
mirror, while the camera body should be centered on and
perpendicular to the telescope axis, with the top and bottom
sides of the camera parallel to the ground. Ray tracing
computations have shown that the spot size increases by about
10\,\% when the longitudinal distance from the center of
curvature changes by $\pm 2$~mm. Therefore, the  accuracy on the
longitudinal position of the pixels on the focal surface should
be better than $\pm 2$~mm.

The intrinsic accuracy of the rigid metal frame of 
of the camera body is very good, at the level of $\pm
0.1$~mm. 
The point-to-point internal accuracy of the pixels on the camera
body is at the level of $\pm 1$ mm in both the longitudinal and transverse
directions, due to the positioning of the photomultipliers and of the 
light concentrators on the rigid frame of the camera body.


The fluorescence buildings are surveyed by standard topographic methods and for each bay two accurately determined reference points are marked on the floor defining a line which corresponds to the azimuth of the telescope axis. The center of the camera is placed in the vertical plane containing this line and at the right nominal height. The camera is then aligned horizontally using a digital
level-inclinometer with a precision better than 0.05$^\circ$ which is placed on the top of the camera body.  Once the center of the camera is correctly positioned and the camera is horizontal, the right orientation is achieved measuring the distance of the four corners of the camera body to the mechanical reference point which is located at the mirror center. These measurements are done with a commercial laser distance-meter with a precision of $\pm 1$ mm mounted on the mirror center.  The alignment procedure should provide a positioning with space accuracy at the level of $\pm 1$mm and angular precision at the level of one millirad i.e. better that 0.1$^{\circ}$.

Measurements of the image of bright stars on the camera focal
surface have verified that the alignment procedure for mirror
elements and camera body meet the design specifications
\cite{stars}. The precision of the absolute pointing of
the telescopes has been checked to within an accuracy of 0.1
$^{\circ}$.

\subsection{The pixel array}\label{2.3}

The hexagonal photomultiplier tube (PMT), model XP3062 
manufactured by Photonis  \cite{photonis} is used to instrument the camera. 
Although their hexagonal shape represents
the best approximation to the pixel geometry, a significant
amount of insensitive area is nevertheless present between the photocathodes.
In fact, some space between
the PMTs is needed for safe mechanical packaging on the focal
surface; moreover, the effective cathode area is smaller than the
area delimited by the PMT glass envelope.  To maximize
light collection and guarantee a sharp transition between
adjacent pixels, the hexagonal PMTs are complemented by light
collectors.

\begin{figure}[htb]
\begin{center}
\includegraphics[width=0.8\textwidth]{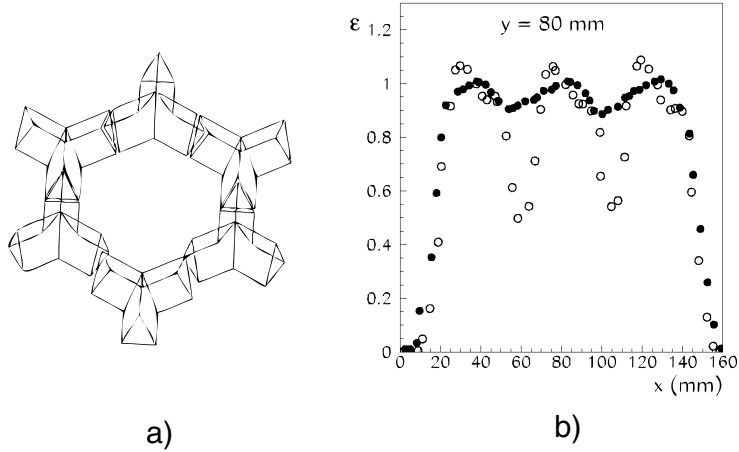}
\caption{a) Six Mercedes stars positioned in order to form a 
pixel. Each Mercedes star has three arms oriented 120$^\circ$
apart. In the drawing the bottom Mercedes is slightly displaced
for clarity. b) Measurement of the light collection efficiency,
with a light spot moved along a line passing over three adjacent
pixels. The full (open) dots represent the measurements performed with (without) Mercedes stars.}
\label{figcamera3}
\end{center}
\end{figure}

The basic element for the pixel light collector is a Mercedes
star, with three arms oriented 120$^\circ$ apart, which is
positioned on each pixel vertex. Six Mercedes stars collect the
light for a given pixel. The geometrical structure of the light
collector for one pixel is shown in Fig.\ref{figcamera3}a.

The Mercedes stars are  made of plastic material covered by
aluminized Mylar{\texttrademark} foils. The arm length is
approximately half of the pixel side length. The arm section is
an equilateral triangle. The base length of 9.2 mm is designed to
cover the insensitive space due to photocathode inefficiency
($\approx$2 mm for each adjacent PMT) plus the maximum space
between the glass sides of the PMTs (of the order of 5 mm). The
triangle height is 18 mm, and the corresponding angle at the
vertex is $14.3^\circ$. 

Each Mercedes star is held by a bar, about 10 cm long, which is
inserted into a 3.2 mm hole located on the position of the pixel
vertices and kept in place by a small O-ring. The geometry of the
light collectors was designed on the basis of the optical system
properties. The range of angles of incidence for the rays on the
camera is in the interval between approximately 10$^\circ$ to
35$^\circ$, as determined by the shadow of the camera and the
aperture of the diaphragm. 
Note that the pixels are defined on the focal surface at 1743 mm from the center of curvature
where the top edges of the Mercedes divide one pixel from its neighbors. 
The PMT cathodes are recessed behind that focal surface by 18 mm.

Monte Carlo ray tracing has shown that the light collection
efficiency, averaged over the camera, is 94\%, assuming a
reflectivity of 85\% for the aluminized Mylar foils of the light
collector surface. Without the light collectors, the collection
efficiency decreases to 70\%. 

The Monte Carlo simulation was experimentally checked using a specially designed light diffusing cylinder with an exit hole for light rays having the same size as the spot produced by the telescope optics. The distribution of the angles of the light rays from this optical device is similar to that produced by the telescopes.
Measurements were made with and without
the light collectors. Results of a scan moving the light spot
over the camera are shown in Fig.~\ref{figcamera3}b. Without light
collectors, a significant loss of light at the pixel borders was
observed. When installed, the light collectors efficiently
recuperate the light loss. From these measurements, the light
collection efficiency averaged over the camera focal surface was
found to be 93\%. 

\begin{figure}[bth]
\begin{center}
\epsfig{file=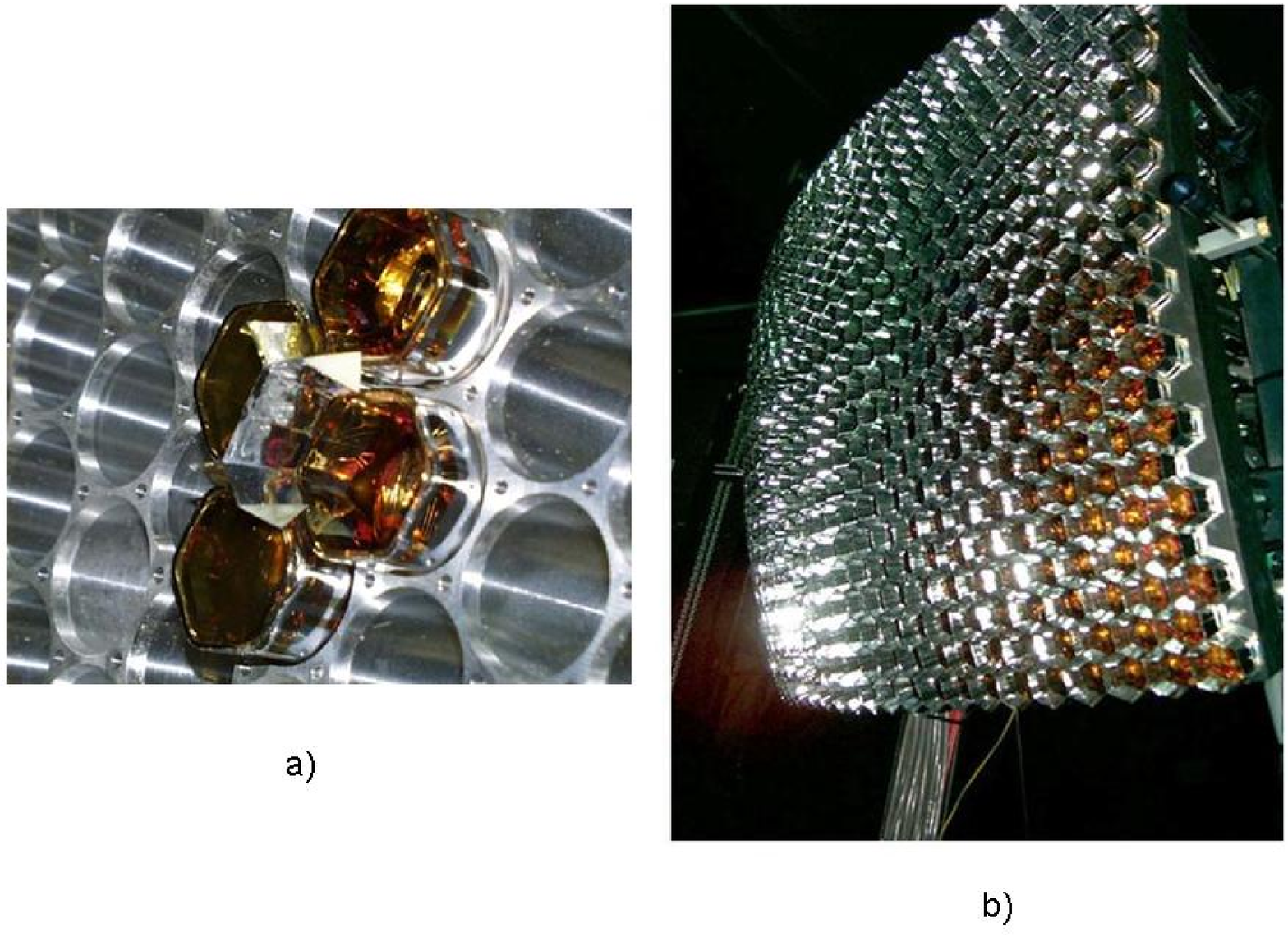,height=12.cm}
\vspace*{-1.6cm}
\caption{a) Detail of the camera body with four PMTs mounted together with two Mercedes stars. The large holes to insert the PMTs and the small ones to mount the Mercedes are visible.
b) Picture of a camera completely assembled with all PMTs and light collectors in place.}
\label{figcamera4}
\end{center}
\end{figure}
            
A picture of a small section of the camera and a full picture of
the camera completely assembled are shown in
Fig.~\ref{figcamera4}.

\subsection{The photomultiplier unit}

The XP3062 photomultiplier \cite{photonis} is an 8-stage PMT with
a hexagonal window (40 mm side to side). It is manufactured with
a standard bialkaline photocathode with quantum efficiency of
about 25\,\% in the wavelength range 350-400~nm. The nominal gain
for standard operation of the FD is set to $5\times10^4$.

\begin{figure}[htb]
\begin{center}
\epsfig{file=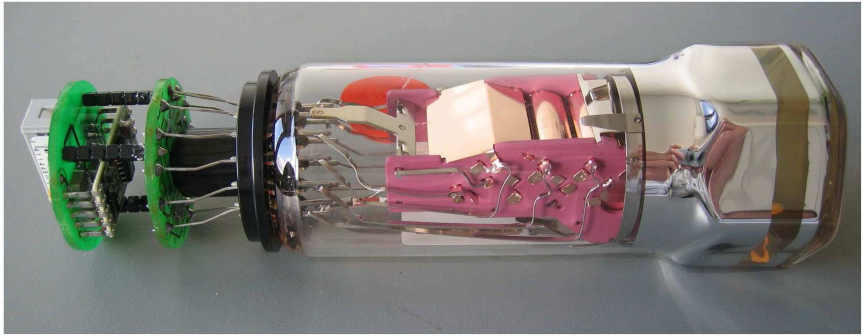,height=5.cm}
\caption{Picture of a PMT unit.}
\label{figcamera5}
\end{center}
\end{figure}

The PMT high voltage (HV) is provided by a HV-divider chain which
forms a single physical unit together with the signal driver
circuitry. This unit, called head electronics (HE), consists of
three coaxially interconnected printed circuits boards (PCBs):
the bias PCB (innermost one), a laser-trimmed hybrid technology
driver circuit (intermediate), and the interface PCB (outermost).
The innermost and outermost PCBs are two-sided and of circular
shape (32\,mm diameter), and are interconnected using high
reliability pin connectors. The HE units were manufactured by
Intratec-Elbau (Berlin, Germany) \cite{intratec} and are soldered
to the flying leads of the PMT. To ensure central mounting of the
HE with respect to the symmetry axis of the PMT, and to improve
mechanical rigidity, a specially-designed plastic structure has
been introduced between the glass tube and a central guidance
hole left in the innermost PCB of the HE. A PMT with the attached
head electronics is shown in Fig.~\ref{figcamera5}.

The HV divider keeps the PMT photocathode at ground and supplies
positive HV to the anode. To dissipate minimum power in the dense
package of HE units, the bleeder is operated at currents of less
than 170\,$\mu$A. Stabilization of the HV potential for large
pulses and in presence of a relatively strong light background is
thus realized by employing an active network \cite{divider} that
uses bipolar transistors in the last three stages of the PMT.
This is necessary for a fluorescence telescope, which is exposed
to the dark sky background and in some cases also to a fraction
of moon light. The normal dark sky background induces an anode
current of about 0.8 $\mu$A.  The active divider ensures that the
gain shift due to the divider chain is less than 1\% for anode
currents up to about 10 $\mu$A.

The driver located on the intermediate ceramics board of the HE receives the
AC-coupled anode signal through a differential input. The anode pulse flows
through a load resistor and reaches one leg of the differential input of the
line driver, while the other leg only picks up the common-mode noise. A
common-mode rejection ratio of 28 dB is obtained between $1-100$~kHz.
The integrated laser-trimmed hybrid circuit supplies a balanced output matching
the 120~$\Omega$ characteristic impedance of the twisted pair cable connecting
the HE to the front-end analog board of the readout electronics (see section
\ref{subsubsec:analog-board}).  

Extensive qualification and acceptance tests \cite{testpmt}  were
performed on the full PMT unit, i.e.\ the PMT with the HE
attached, using automatic test systems. Among the tests were
measurements of the HV dependence of the gain, which are needed
for a proper grouping of the PMTs with similar gain. The relative
photocathode sensitivity at different wavelengths, the linearity
of the PMT unit, and the photocathode uniformity were also
measured. Before and after installation, the relevant information
for each individual PMT unit is collected into a database which
tracks the PMT characteristics, including calibration
measurements performed {\it in situ}.


To reduce the cost of the power supplies, the photomultipliers of each camera
are organized into ten groups of 44 units.  Each group has similar gain
characteristics, and is powered by a single HV channel. The spread of the gains
within a group is about $\pm$ 10\%.

Cables are distributed to the PMT electronics through distribution boards 
positioned behind the camera and within its shadow, i.e., without
causing additional obscuration. These boards serve groups of 44
PMTs of the camera,  supplying HV and LV and receiving the
differential signals from the drivers in twisted pair cables.  On
the board a fuse for overcurrent protection of each LV line is
provided. From the boards, round shielded cables carry the PMT
signals on twisted pair wires to the front-end crate, located at
the base of the camera support.



\section {Electronics and data acquisition system}\label{electronicsystem}

The FD telescopes are used to record fluorescence signals of widely varying
intensity atop a sizeable, and continuously changing, light background.  This
presents a significant challenge for the design of the electronics and data
acquisition system (DAQ), which must provide a large dynamic range and strong
background rejection, 
while accepting any physically plausible air shower. 
The
DAQ must also allow for the robust, low-cost, remote operation of the FD
telescopes.  And finally, the absolute FD-SD timing offset must be sufficiently
accurate to enable reliable hybrid reconstruction.

\begin{figure}[ht]
  \begin{center}
    \includegraphics[width=0.8\textwidth]{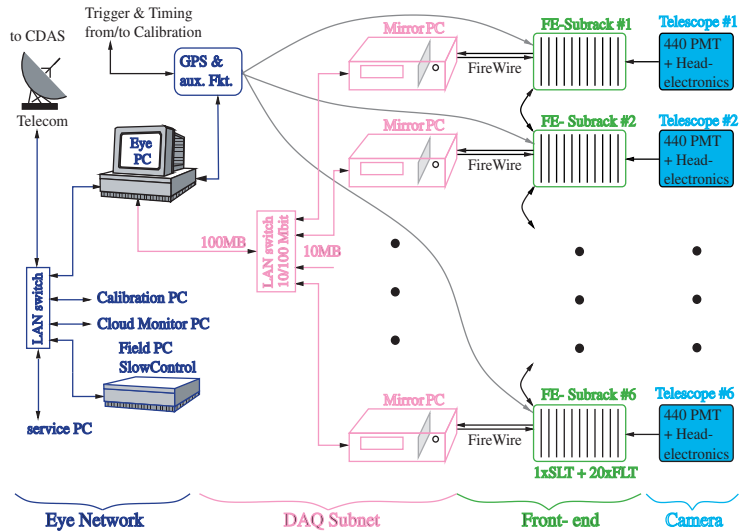}
    \caption{Readout scheme of an FD site 
             The flow of recorded data is 
             right to left.}
    \label{sysoverview}
  \end{center}
\end{figure}

The FD electronics are responsible for anti-alias filtering, digitizing, and storing signals
from the PMTs.  As the PMT data are processed, they are passed through a
flexible three-stage trigger system implemented in firmware and software.  The
remaining high-quality shower candidates are packaged by an event builder for
offline shower reconstruction.  For each shower candidate, a hybrid trigger is
generated for the surface detector.  An overview of the complete trigger
sequence is presented in Table~\ref{table:fd-triggers}.

The organization of the electronics and DAQ is hierarchical, reflecting the
physical layout of the FD buildings.  Figure~\ref{sysoverview} shows the
readout scheme of one FD site, 
 divided into four logical units: head
electronics for $440$~PMTs~$\times$~6~telescopes, which provide low and high
voltage; front-end (FE) sub-racks, where the signals are shaped and digitized,
and where threshold and geometry triggers are generated; the DAQ subnet, in
which six telescope PCs (MirrorPCs) read out the stored data and perform
additional software-based background rejection; and an FD site network, 
 in which a
single EyePC merges triggers from the six telescopes and transfers them to the
Observatory Central Data Acquisition System (CDAS) in Malarg\"{u}e.
The FD site 
network also contains a Slow Control PC to allow for remote operation of the
building.  The camera electronics and PCs are synchronized by a clock module
based on the Motorola Oncore UT+ GPS receiver, the same receiver used in the SD
array.

\begin{table}[tbh]
  \begin{center}
  \def\arraystretch{1.05}
  \begin{tabular}{|c|c|l|l|}\hline
    \multicolumn{4}{|c|}{
      \rule[-3mm]{0mm}{8mm}\bfseries FD Trigger Sequence} \\
      \hline
    \slshape Trigger Level &
    \slshape Location & 
    \slshape Purpose &
    \slshape Event Rate \\
    \hline
    1 &
    FE sub-racks & 
    pixel threshold & 
    \multirow{2}{*}{100~Hz~pixel$^{-1}$} \\
    (FLT) & (FLT boards) & trigger & \\
    \hline
    2 &
    FE sub-racks & 
    track shape & 
    \multirow{2}{*}{$0.1-10$~Hz~telescope$^{-1}$} \\ 
    (SLT) & (SLT board) & identification & \\
    \hline
    3 &
    MirrorPCs & 
    lightning & 
    \multirow{2}{*}{$0.01$~Hz~telescope$^{-1}$} \\
    (TLT) & (software) & rejection & \\
    \hline
    \multirow{2}{*}{T3} &
    EyePC & 
    event builder, & 
    \multirow{2}{*}{$0.02$~Hz~building$^{-1}$} \\
    & (software) & hybrid trigger & \\
    \hline
  \end{tabular}
  \caption{Trigger sequence for FD events.  At each telescope, events are
           selected based on channel thresholds (FLT), track shape (SLT), and
           lightning rejection (TLT).  Events passing the TLT are merged by an
           event builder on the FD EyePC.  If the event passes further quality
           cuts for a simple reconstruction, a hybrid trigger (T3) is sent to
           CDAS.}
  \label{table:fd-triggers}
  \def\arraystretch{1.5}
  \end{center}
\end{table}

\subsection{Front-end electronics}\label{subsec:fe-electronics}

Each FD camera is read out by one front-end sub-rack and an associated
MirrorPC.  The front-end electronics contain 20 Analog Boards (ABs), and
each AB receives data from a column of 22 PMT channels.  The boards are
designed to handle the large dynamic range required for air fluorescence
measurements; at the energies of interest for the Pierre Auger Observatory,
this means a range of 15~bits 
and 100~ns timing~\cite{gemmeke}.  The sub-racks also contain dedicated boards for
hardware triggers: 20 First Level Trigger (FLT) boards for pixel triggers,
and one Second Level Trigger (SLT) board for track identification within the
camera image.

\subsubsection{The Analog Board}\label{subsubsec:analog-board}

\begin{figure}[ht]
  \begin{center}
    \includegraphics[width=0.8\textwidth]{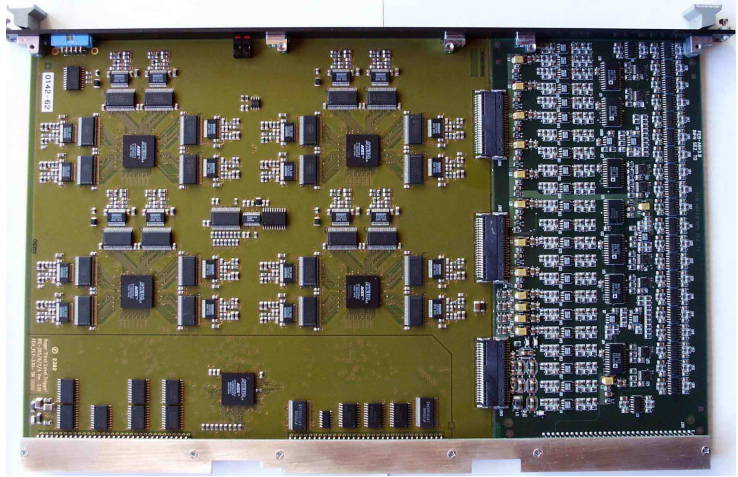}
    \caption{Photo of the FLT (left) and Analog Board (right): Both boards are
             connected by three 50-pin SMC connectors.  A stiffener bar and
             common front panel provide mechanical rigidity.}
    \label{fig:FLTAB}
  \end{center}
\end{figure}

The purpose of the AB is to receive inputs from the head electronics on the PMT
camera and adapt them for digitization by the analog-to-digital converters
(ADCs) located on a corresponding FLT board.  As shown in Fig.~\ref{fig:FLTAB},
the analog and FLT boards are physically connected by three 50-pin SMC
connectors.  The combined front-end module measures 367~mm$\times$220~mm, and
is housed in a 9U standard crate next to each FD telescope.  Every crate
contains 20 modules in total.

One AB channel comprises the following logical blocks: receiver, gain stage,
anti-aliasing filter, and dynamic range adapter.  The channel receives input
from the HE on the PMT camera via a distribution board.  Individual pixel
enabling is performed by a fast analog switch, which is also used
to generate an internal test pulse upon request from the FLT logic.  In each
channel, the AB:
\begin{itemize}
  \item performs a differential to single-ended conversion of the signal;
  \item adjusts the channel gain;
  \item applies an anti-aliasing filter before signal sampling;
  \item adapts the 15-bit dynamic range to the 12-bit ADCs;
  \item provides an injection point for test pulses.
\end{itemize}

%
%

The telescope PMTs are coarsely gain-matched during the installation of the PMT
camera to avoid expensive individual HV control.  Additional fine-tuning of the
\textsl{channel} gains is achieved by means of digital potentiometers connected
in series with a resistor on the feedback loop of each non-inverting gain
stage.  The devices can change individual channel gains up to a factor of about
$1.9$, and allow for gain matching of the channels in the camera to within
0.6\%.

Prior to sampling, the PMT signal is processed by an anti-aliasing filter to
match the 10~MHz digitization rate.  A fourth-order Bessel filter with a cutoff
frequency of 3.4~MHz has been implemented in the AB as a compromise between
reconstruction error and circuit complexity.  The Bessel filter scheme,
featuring a linear dependence of the transfer function phase with frequency,
was selected after a detailed study of optimal filters to minimize distortion
of the current signal shape.

\begin{figure}[tbh]
  \begin{center}
    \includegraphics [width=1.0\textwidth]{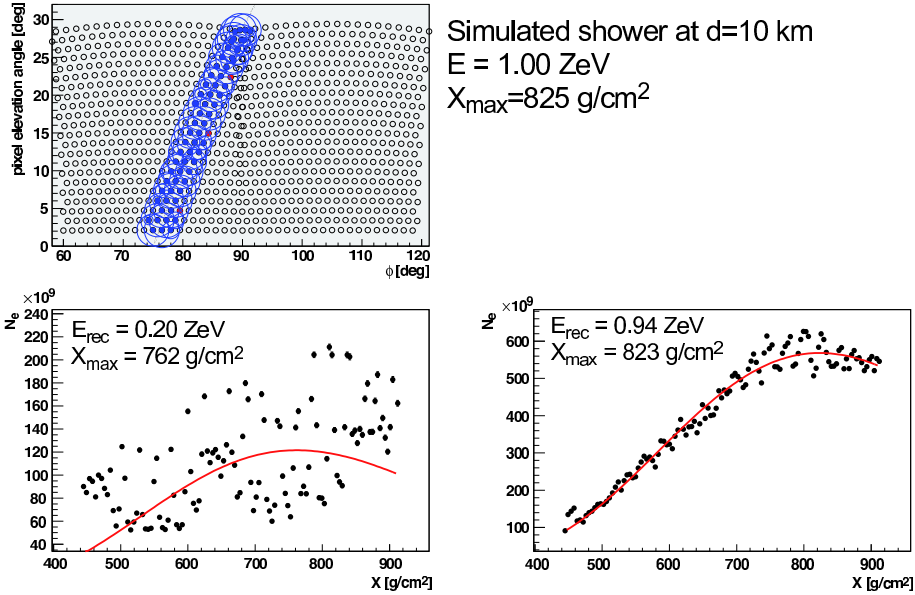}
    \caption{
    Simulated air shower of $10^{21}$~eV at 10 km from the telescope.  The
    top-left figure shows the light track in the cameras.  The lower left
    figure shows the reconstructed shower profile without use of the virtual
    channel information. Making use of the virtual channel (lower right),
    the shower parameters can be reconstructed with minimal systematic
    distortions.}
    \label{fig:virtual-channel}
  \end{center}
\end{figure}

The final component of an AB channel is the dynamic range adapter.  The FD
records signal sizes between 3 and $10^5$ photoelectrons per 100~ns, covering a
dynamic range of $15$~bits.  Rather than digitize the full range of signal
sizes, an optimal cost-effective solution using dynamic range compression to
12-bit ADCs has been implemented in the analog electronics design.

The compression technique, which uses \textsl{virtual channels}, relies on the
fact that the the shower signal does not appear on all pixels simultaneously;
instead, each pixel is triggered in a well-ordered time sequence.  Every
channel is configured with a high and low gain of about $20$ and $1$,
respectively.  The high gain is optimized for the most frequent small- and
intermediate-amplitude pulses, and is digitized pixel by pixel.  The signal of
the low gain stage is routed together with signals from 10 other 
non adjacent pixels in
an analog summing stage.  The sum signal is then processed by a virtual channel
with a gain near unity.

When a nearby high-energy shower is detected by the FD, typically only one channel out of
the group of 11 will saturate its 12-bit range at a given time.  Even in case of multiple 
saturation in the same group, this will never occur in overlapping time bins. Therefore, the
signal can be recovered from the virtual channel without ambiguity.
This is
demonstrated in Fig.~\ref{fig:virtual-channel}, which depicts a simulated
$10^{21}$~eV shower with a landing point 10~km from the FD telescopes.  The
high-gain channels saturate during the development of the shower, and an
attempt to reconstruct the shower using only these channels leads to
significant systematic distortions in the shower profile.  However, with the
information in the virtual channel, the shower is reconstructed with almost no
systematic biases\cite{vivi}.

\subsubsection{First Level Trigger (FLT) Module}\label{subsubsec:flt}

The First Level Trigger (FLT) module is the heart of the digital front-end
electronics.  The module processes the data from one 22-channel column.  Its
main tasks are to:
\begin{itemize}
  \item continuously digitize signals from the AB at 10~MHz;
  \item store the digitized raw data in memory for later readout;
  \item measure the pixel trigger rate for each channel;
  \item compensate for changing background conditions and maintain a pixel 
        trigger rate of 100 Hz by adjusting the pixel trigger thresholds;
  \item allow access to raw data memory and internal registers;
  \item provide a digital interface to the AB to generate test pulses and to set
        the analog gain at the AB;
  \item calculate the baseline offset and its fluctuation averaged over a
        6.5~ms period;
  \item calculate the multiplicity (number of triggered pixels) in one
        column.
\end{itemize}

The functions of the FLT are implemented in FPGA (Field Programmable Gate Array) 
firmware to improve the
flexibility and cost-effectiveness of the module.  A pipelined 12-bit ADC
(ADS804) is used to digitize the signal at 10~MHz in each channel, and the data
are stored with a 4-bit status word in 64k$\times$16-bit SRAMs.  The address
space of each SRAM is divided into 64 pages of 1000 words.  In the absence of
triggers from the SLT module (described in the next section), each page works
as a circular buffer to hold the ADC values of the previous 100~$\mu$s.  When
an SLT trigger occurs, all FLT boards synchronously switch to the next unused
memory page, whose address is provided by the SLT.

\begin{figure}[ht]
  \begin{center}
    \includegraphics[width=0.8\textwidth]{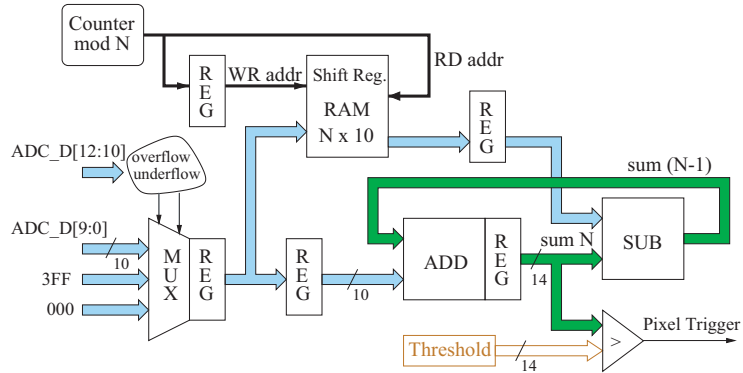}
    \caption{Generation of the First Level Trigger in one channel.  A moving sum
             of 5 to 16 values is compared to an adjustable threshold.  The use
             of the sum improves the S/N-ratio, and allows for regulation of
             the trigger rate.}
    \label{fig:boxcar}
  \end{center}
\end{figure}

The main task of the FPGA logic is to generate the pixel trigger (FLT) using a
threshold cut on the integrated ADC signal.  The FLT is shown schematically in
Fig.~\ref{fig:boxcar}.  A moving ``boxcar'' sum of the last $n$ ADC samples
is compared to a 14-bit threshold. Here $n$ is a fixed number of time bins 
which can be chosen in the range $5 \leq n \leq 16$.  The threshold is
dynamically adjusted to maintain a pixel trigger rate of 100
Hz.
When the sum exceeds the threshold, a pixel trigger is generated.  The use of
the sum substantially increases the signal-to-noise (S/N) ratio for each
channel; for $n=16$, S/N improves by a factor of 2.8~\cite{ICRC01}.  When the
moving sum drops below the threshold, a retriggerable mono-flop extends the
pixel trigger for an adjustable period of 5~$\mu$s to 30~$\mu$s common to all
pixels, increasing the chance of coincident pixel triggers.

%

The rate of pixel triggers, called the ``hit rate,'' is measured in parallel by
counters for each channel. It is used to adjust the threshold in such a manner
that the hit rate is kept constant at 100~Hz under variable background light
conditions.  The background light levels seen by each PMT can also be monitored
by analysis of the variances of the ADC values. Therefore, the FPGAs regularly
evaluate the mean and variance of the channel baselines every 6.5~ms using
65536 consecutive ADC samples~\cite{IEEE02}.  The measured channel hit rates,
thresholds, offsets, and variances are periodically queried and stored in a
monitoring database for experimental control (see
section~\ref{sec:monitoring}).


Finally, the multiplicity, or the number of pixels triggered simultaneously
within 100 ns, is calculated for each FLT and for the full camera.  The number
of channels above threshold in one 22-pixel column is tracked on the FLT board.
The total sum for the full camera is evaluated at the SLT board using
daisy-chained lines on the backplane.  The chronological sequence of
multiplicity values carries information about the temporal development of the
camera image, which is used by the software trigger (see
section~\ref{subsubsec:TLT}).

\subsubsection{The Second Level Trigger (SLT)}\label{subsubsec:slt}

\begin{figure}[tbh]
  \begin{center}
    \vspace{0.4 cm}
    \includegraphics[width=0.8\textwidth]{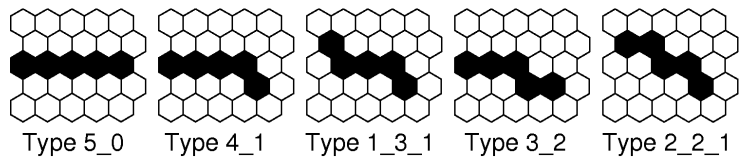}
    \caption{Fundamental types of pattern regarded as straight track segments.}
    \label{fig:slt-pattern}
  \end{center}
\end{figure}

The pixel triggers generated for each channel in the 20 FLT boards of the
FE sub-rack are read out by a Second Level Trigger (SLT) board.  The functions
of the SLT are implemented in FPGA logic, and its primary task is to generate
an internal trigger if the pattern of triggered pixels follows a straight
track \cite{szadkowski}.  The algorithm searches for track segments at least five pixels in
length.  It uses the fundamental patterns shown in Fig.~\ref{fig:slt-pattern},
as well as those created by rotations and mirror reflections of these segments.

During data acquisition, some tracks will not pass through every pixel center,
and therefore some PMTs along the track may not record enough light to trigger.
To allow for this situation, and to be fault-tolerant against defective PMTs,
the algorithm requires only four triggered pixels out of five. Counting all
different four-fold patterns originating from the five-pixel track segments in
Fig.~\ref{fig:slt-pattern}, one finds 108 different pattern classes.

A brute-force check of the full $22\times20$ camera pixel matrix for the 108
pattern classes would result in 37163 combinations in total.  Therefore, the
SLT uses a pipelined mechanism that scans for tracks on a smaller $22\times5$
sub-matrix of the camera.  Every 100~ns, the FLT pixel triggers of two adjacent
columns ($22\times2$ pixels) are read into a pipelined memory structure.
During the scan of the full camera, coincidence logic identifies the pattern
classes 1 to 108 and generates a trigger if a pattern is found.  The full scan
lasts 1~$\mu$s, and the pixel trigger information scanned during this time is
stored in the \textsl{pixel memory} of the SLT, which is organized as a ring
buffer for 64 events (as on the FLT).  The FLT multiplicities are also stored
in the same memory, providing complementary measures of the spatial and
temporal size of the event.


Note that in addition to generating the geometrical trigger, the SLT board also
supplies bookkeeping information for the FE sub-rack.  In particular, the board
provides a time stamp for each event using synchronization signals sent by the
GPS module over a dedicated line.

\subsubsection{Test Systems}\label{subsubsec:fe-tests}


The large number of channels in each telescope and the complexity of the
electronics and triggers require the use of an automatic testing system.  The
testing and calibration of the full optics-PMTs-electronics chain is described
in section~\ref{calibration}.  However, the system also allows for tests of the
electronics alone.  A test-pattern generator has been implemented on the
front-end boards to create pulses of variable width and amplitude at the input
of each analog channel.  These test pulses are used to check the analog and
digital functions of the FLT and SLT boards.  The system is independent of the
full camera, and can be used for maintenance when the HV/LV supply of the
camera is turned off.

\subsection{DAQ Software and Software Trigger}

Once an event has been processed by the DAQ hardware and stored in FLT and SLT
memory, it can be read out and analyzed by trigger software in the MirrorPCs.
Each MirrorPC is connected to the digital boards in the FE sub-rack via a
FireWire interface~\cite{firewire}, and the MirrorPCs communicate with the
EyePC through a 10/100 Mbit LAN switch.

The DAQ system can handle data from three types of sources:
\begin{enumerate}
  \item external triggers;
  \item calibration events and test pulses;
  \item real air shower events.
\end{enumerate}
External triggers are primarily artificial light sources used for atmospheric
monitoring, such as laser shots from the Central Laser Facility~\cite{CLF}.
The various types of calibration pulses are described in
section~\ref{calibration}.

Events triggered externally, or by the calibration and test-pulser systems,
receive special trigger bits in the SLT status word to distinguish them from
real air shower data.  When the MirrorPCs read out such events from the FE
modules, they are sent directly to the EyePC for event building and storage in
dedicated raw data files.  In contrast, when the MirrorPCs read out true air
shower data, the events are processed by a Third Level Trigger.  The
surviving data are sent to the EyePC, which builds an event from the coincident
data in all telescopes and generates a hybrid trigger (T3) for the surface
array.

%
%
%

\subsubsection{The Third Level Trigger (TLT)}\label{subsubsec:TLT}

The Third Level Trigger (TLT) is a software algorithm designed to clean the air
shower data stream of noise events that survive the low-level hardware
triggers.  It is optimized for the fast rejection of triggers caused by
lightning, triggers caused by muon impacts on the camera (see
Fig.~\ref{figcamera6}), and randomly triggered pixels.

During the austral summer, distant lightning can significantly disrupt the
normal data acquisition of the FD telescopes.  In good atmospheric conditions,
the SLT will detect one to two events per minute per telescope.  However,
lightning can cause large parts of the camera (i.e., hundreds of pixels) to
trigger in bursts of several tens of events per second. 
If not filtered, the bursts will congest
the 64-event buffers in the FLT and SLT boards, and the telescope readout
(including ADC traces) will suffer a substantial increase in dead time.

The TLT is designed to efficiently filter lightning events without performing
costly readouts of the complete ADC traces.  The algorithm achieves high speed
by reading out only the FLT multiplicities and the total number of triggered
pixels.  Using several thousand true showers and background events from one
year of data, a dedicated study was performed to determine efficient lightning
cuts based on the time development of the multiplicity  and the
integral of the multiplicity over the whole event consisting of 1000
subsequent multiplicity values~\cite{NewTLT}.  With these cuts,
approximately 99\% of all 
lightning events are rejected by the TLT during a short (50~$\mu$s) decision
time.

The multiplicity-based lightning rejection effectively removes noise events
with more than 25 pixels.  The remaining  events with smaller number
of triggered pixels, which are contaminated
by random pixel triggers and muon impacts, are filtered in a second step.  The
correlation between the spatial arrangement and peak signal times of triggered
pixels is used to discard noisy channels far off the light track~\cite{NewTLT}.
The full ADC traces must be used to determine the pixel trigger times, but this
readout does not appreciably increase the telescope dead time when the number
of pixels is small.

The TLT performance has been validated with simulated showers and data recorded
during different weather and sky background conditions.  Although the exact
behavior of the algorithm depends on the actual conditions, approximately
$94$\% of all background events are correctly identified by the TLT.  The
fraction of true showers rejected by the trigger is below 0.7\%.

\subsubsection{Hybrid Trigger (T3)}\label{subsubsec:hybridT3}

Events passing the TLT in each telescope are sent to the EyePC through the DAQ
subnet.  An event builder running on the EyePC merges coincident events from
adjacent telescopes.  The EyePC also sends a hybrid trigger, called a T3, to
the CDAS in Malarg\"{u}e.

The T3 acts as an external trigger for the surface array.  Its purpose is to
record hybrid events at low energies (below $3\times10^{18}$~eV) where the
array is not fully efficient and would not often generate an independent
trigger.  Hybrid events at these energies occur within $20$~km of the FD
buildings and usually do not trigger more than one or two SD stations.
However, as discussed in section~\ref{reconstruction}, even limited SD data are
sufficient for high-quality hybrid reconstruction.

The T3 algorithm is used to calculate a preliminary shower direction and ground
impact time with a simple online reconstruction.  Once these data arrive at the
CDAS, a request is sent to the SD for signals recorded close to the calculated
impact time.  For each T3 trigger, the SD stations nearest the FD building
(comprising approximately one-quarter of the array) are read out.  The FD and
SD data are merged offline for subsequent hybrid analysis.

\section{Calibration}\label{calibration}


\subsection{Introduction}

The reconstruction of air shower longitudinal profiles and the
ability to determine the total energy of a reconstructed shower
depend on being able to convert ADC counts to a light flux at the
telescope aperture for each channel that receives a portion of
the signal from a shower.
It is therefore necessary to have some method for evaluating the
response of each pixel to a given flux of incident photons from
the solid angle covered by that pixel, including the effects of
aperture projection, optical filter transmittance, reflection at
optical surfaces, mirror reflectivity, pixel light collection
efficiency and area, cathode quantum efficiency, PMT gain,
pre-amp and amplifier gains, and digital conversion.  While this
response could be assembled from independently measured
quantities for each of these effects,
an alternative method in which the cumulative effect is measured
in a single end-to-end calibration is employed here.

The absolute calibration of the fluorescence detectors uses a
calibrated 2.5~m diameter light source (known as the ``drum'') at
the telescope aperture, providing same flux of light to each
pixel. The known flux from the light source and the response of
the acquisition system give the required calibration for each
pixel.  In the lab, light source uniformity is studied using CCD
images and the intensity is measured relative to NIST-calibrated
photodiodes~\cite{NIST}. Use of the drum for gain adjustment and calibration
provides a known, uniform response for each pixel in each camera of the FD detector.
The average response of the FD is approximately 5 photons/ADC bin.

Three additional calibration tools are used as well:

\begin{itemize}
\item Before and after each night of data taking a relative calibration of the PMTs is performed. This relative calibration
is used to track both short and long term changes in detector
response.
\item The relative FD response has been measured at wavelengths of 320, 337, 355, 380 and 405~nm, defining a spectral response
curve which has been normalized to the absolute calibration.
\item An independent check of the calibration in some phototubes is performed
using vertical shots from a portable laser in the field.  
\end{itemize}

The sections below describe the hardware and use of these
calibration systems. 

\subsection{Light source and drum} 

\begin{figure}[b]
\begin{center} 
\includegraphics[width=0.6\textwidth]{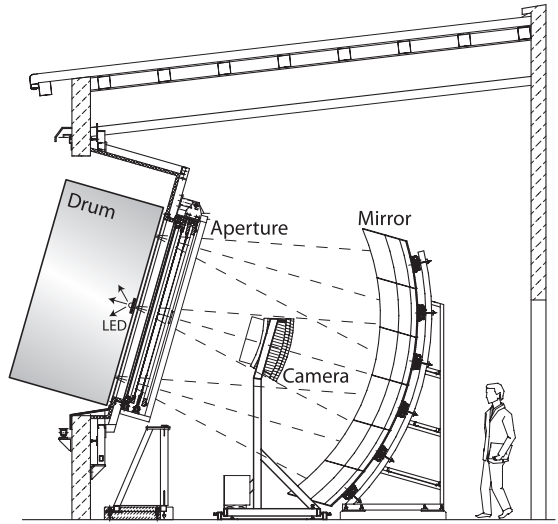}
\end{center}
\caption{A schematic showing the drum mounted in a telescope aperture.}
\label{drum-aperture-schematic}
\end{figure}

The technique~\cite{abscal} is based on a portable 2.5 m
diameter, 1.4~m deep, drum-shaped light source which mounts on
the exterior of the FD apertures (see
Fig.~\ref{drum-aperture-schematic}). The source provides a pulsed
photon flux of known intensity and uniformity across the
aperture, and simultaneously triggers all the pixels in the
camera. The surface of the drum is a good Lambertian.
This means that the light emitted per unit solid angle from
any small area $A$ depends only on the angle $\theta$ with
respect to the normal direction according to $I(\theta)=I_0 A
\cos(\theta)$.  Looking at the disk from angle $\theta$ and
distance $d$ with a fixed solid angle $\omega$, the viewed
disk area is $d \omega/\cos(\theta)$.  So the intensity is
$I(\theta)=I_0 d \omega$.  The intensity is independent of
viewing angle for a perfect Lambertian.  Fig. \ref{drum-unif} shows that
pixels of a CCD camera (each seeing a small portion of the
drum surface) measure the same intensity regardless of the
viewing angle to the drum.

To produce diffuse light inside the drum, illumination is provided by a pulsed UV
LED~\cite{Nichia} ($375\pm12$~nm), mounted against the face of a
2.5 cm diameter $\times$ 2.5~cm long Teflon{\texttrademark}
cylinder. The Teflon cylinder is mounted in a 15 cm diameter
reflector cup, which is mounted flush to the center of the drum
front surface, illuminating the interior and the back surface of the drum.  
The LED is inserted down the axis of the drum from the back through a pipe. 
A silicon detector attached to the opposite end of the teflon
cylinder monitors the light output for each pulse of the
LED.  

The drum was constructed in sections, using laminations of
honeycomb core and aluminum sheet.  The sides and back surfaces
of the drum interior are lined with Tyvek\texttrademark, a
material diffusively reflective in the UV.  The reflecting
surfaces of the cup are also lined with Tyvek.  The front face of
the drum is a 0.38~mm thick Teflon sheet, which transmits light
diffusively.

\subsection{Calibration of the drum}

The absolute calibration of the drum light source intensity is
based on UV-enhanced silicon photodetectors, calibrated at
NIST to $\pm 1.5$\%. While the small surface area and low response
of these detectors preclude detection of the small photon flux
from the drum surface directly, the photodiode calibration can be
transferred to a more sensitive PMT/DAQ system.

To establish the absolute flux of photons emitted from the drum
surface, a reference PMT is placed on the drum axis, 14~m from
the surface.  The LED light source in the drum is pulsed for a
series of 5~$\mu$s pulses and the charge from the PMT is
integrated and recorded for each pulse, resulting in a histogram
of the distribution of the observed integrated flux.

On an optical bench, the PMT is then exposed to a small diffuse
LED light source with neutral density filter. This filter adjusts 
the light intensity to the level similar to that of the drum. 
The LEDs are pulsed with the same driving circuitry as for
the drum, and the intensity is set to a series of values
producing a series of histograms with PMT centroids surrounding
that from the drum measurement above.  At each of these intensity
settings a second measurement is made in which the PMT is
replaced by the NIST-calibrated photodiode and a neutral density
filter in the source is removed, increasing the intensity to a
level measurable by the photodiode.  For this second measurement,
the LEDs are run in DC mode.  The relationship of PMT response to
photodiode current is found to be very linear.  The flux of
photons at 14~m from the drum surface is calculable from the
active area of the photodiode, the neutral density filter
reduction factor, the LED pulsed/DC duty factor, the NIST
calibration for the photodiode, and the value of photodiode
current corresponding to the PMT-drum centroid. The value of 
photodiode current is interpolated from the linear response-current 
relationship given above. The resulting uncertainty in the drum intensity 
is 6\%.  Additional uncertainty contributions related to use of the 
drum at the FDs \cite{Knapik:2007yd}, such as temperature dependencies, 
along with camera response variations in time and spectral 
characteristics of the LED light source, combine with the 
drum intensity uncertainty resulting in an
overall uncertainty of 9\% for the absolute FD calibration.


\subsection{Drum relative uniformity measurements}

\begin{figure}[ht]
\begin{center} 
\mbox{\epsfig{figure=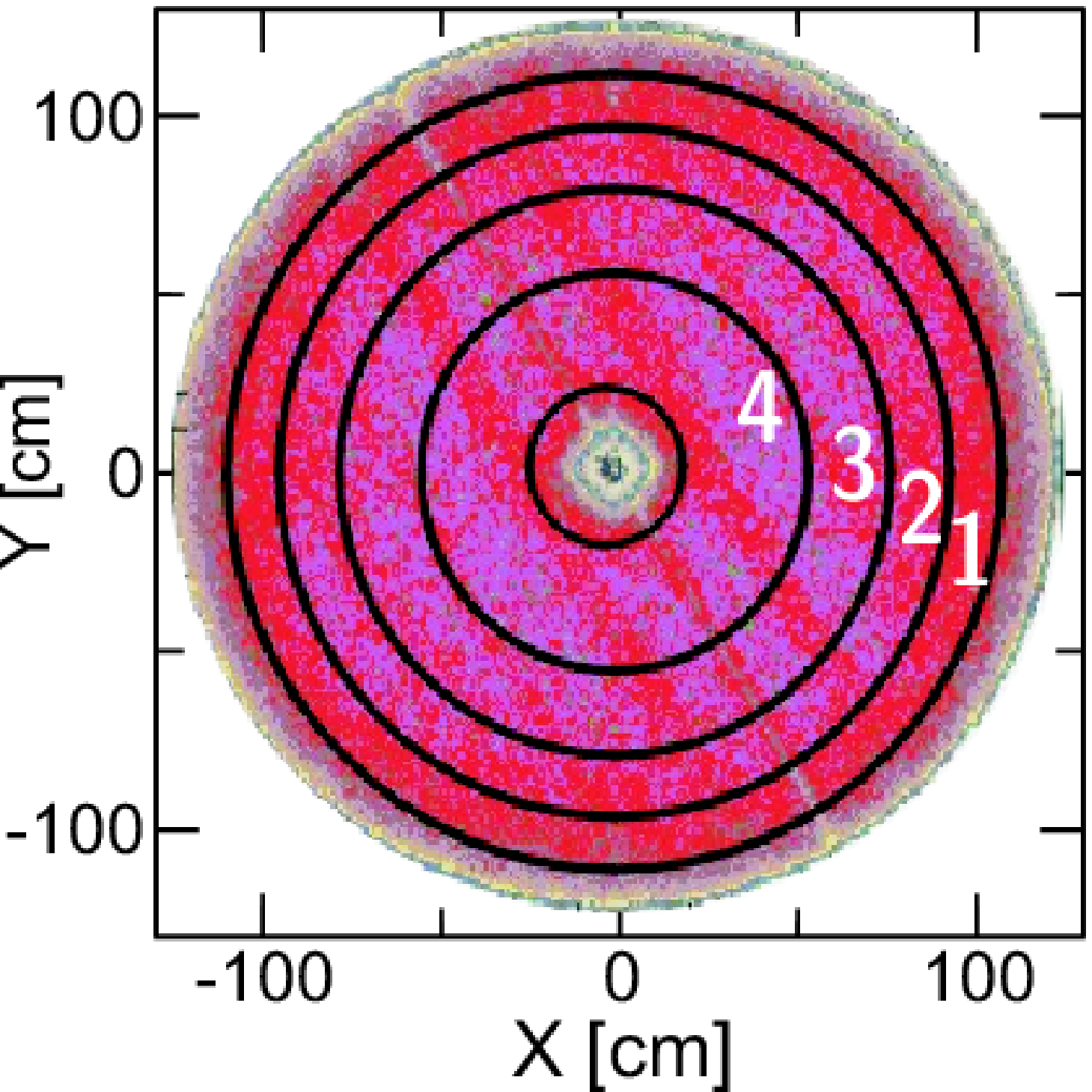,height=3.5cm}
      \epsfig{figure=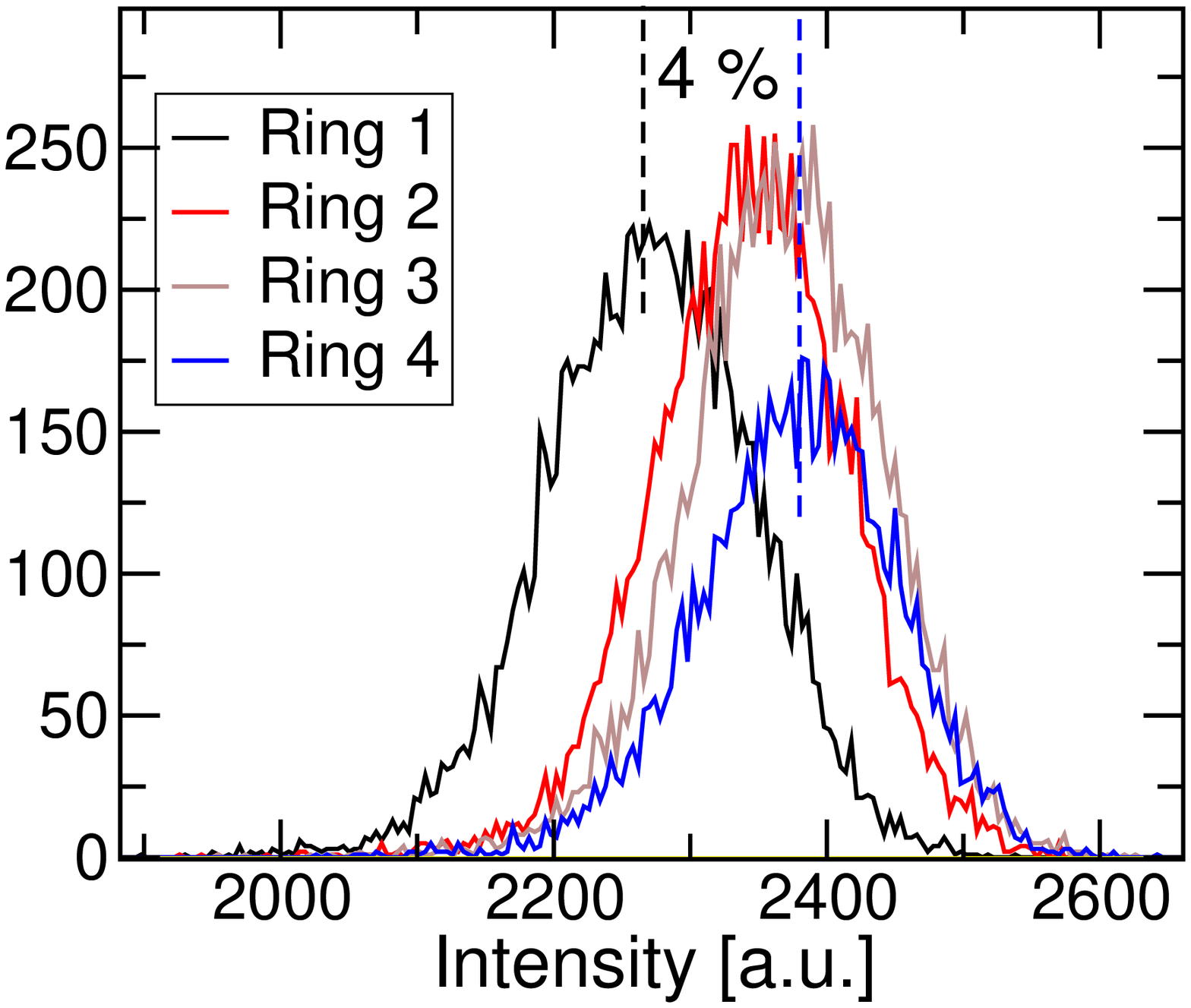,angle=0,height=3.5cm}
      \epsfig{figure=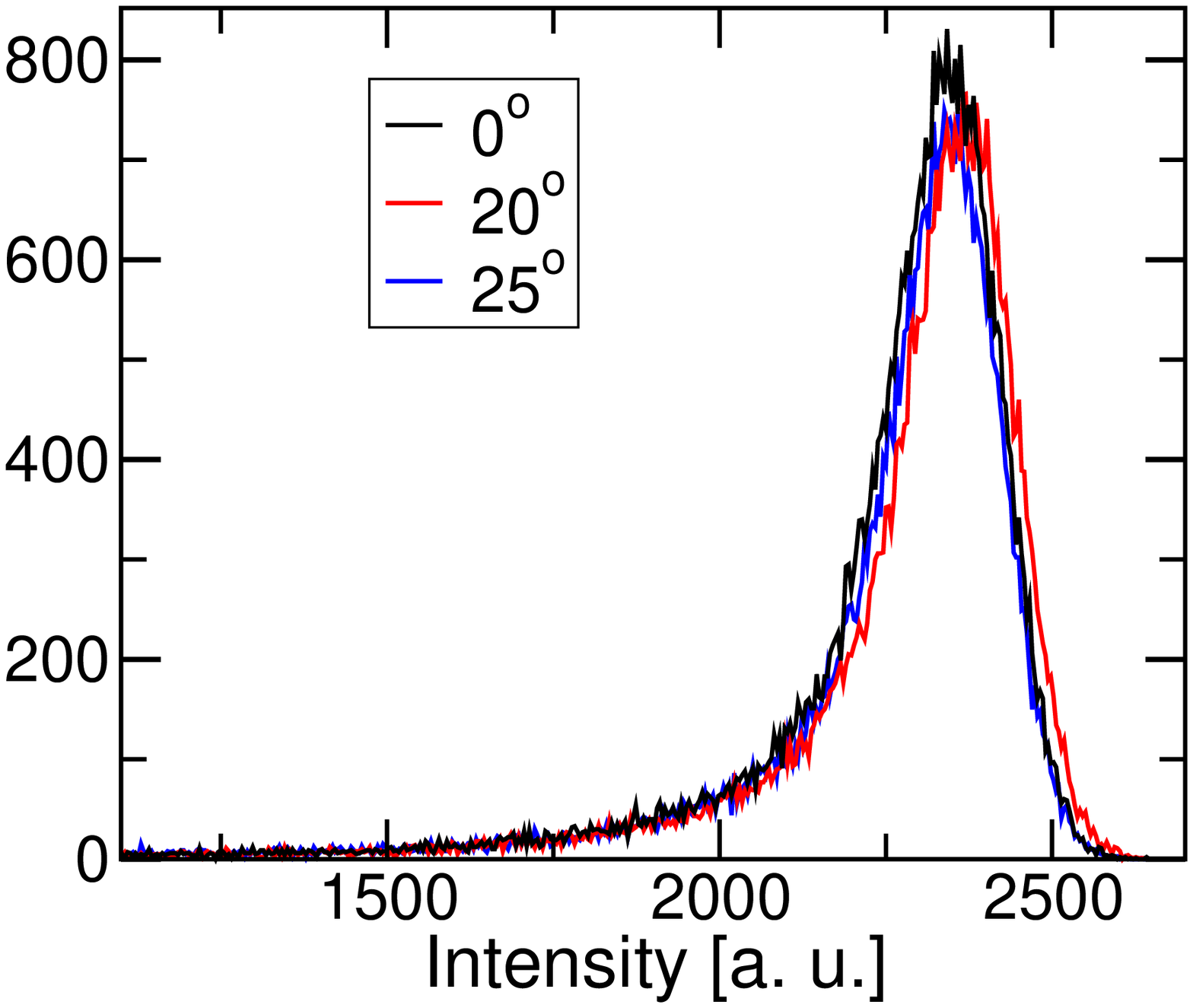,angle=0,height=3.5cm}}
\end{center}
\caption{{\it Left:}
CCD image at 0$^\circ$ drum angle, showing the
defined rings for relative intensity analysis.
Production deformations in the teflon material can be seen.
{\it Middle:} A plot of the observed pixel intensities in the defined
regions of the drum, shown in the left panel; 
{\it Right:} The results of angular relative intensity measurements at 0, 20, and 25$^{\circ}$. 
inclinations of the drum with respect to the CCD camera.} 
\label{drum-unif}
\end{figure}

Uniformity of light emission from the drum surface is important,
since the pixels in a FD camera view the aperture at varying
angles. In addition, for each pixel, a different part of the
aperture is blocked by the camera itself. Studies were made of
uniformity of emission across the surface and as a function of
viewing angle. These uniformity measurements were made using a
CCD, viewing the emitting surface of the drum from a distance of
14~m (see Fig.~\ref{drum-unif}).  Images were recorded with the
drum axis at angles of 0, 10, 20 and 25 degrees relative to the
CCD axis, covering the range of the Auger telescope field of view
($0 - 21^{\circ}$).  For these images, the UV LEDs were powered
continuously.

A 0$^\circ$ image was used to analyze the uniformity of emission
over the drum surface.  Using software, concentric circles were
drawn on the image, defining annular regions of increasing
radius, as shown in the figure. The intensity of the pixels in
each region was analyzed to obtain the intensity as a function of
radius. In the area defined by the 2.2~m aperture radius, the
relative uniformity of intensity is constant over the area to
about $\pm$2\% except for a central dark spot. 
The variation with viewing angle of a section of
the drum image is also shown in Fig.~\ref{drum-unif}.  

The measured drum non-uniformities are small (diagonal stripes
in Fig.~\ref{drum-unif}, intensity decrease with radius, etc.),
indicating that the FD pixels see similar intensities integrated
over the drum surface.  While perfect drum uniformity is
desirable, the present non--uniformities are acceptably small and
well mapped over the surface of the drum.  A ray-tracing program
using the uniformity and angular intensity information from the
CCD images shows less than 1\% variation in total flux seen by
the pixels, and corrections are applied for these variations.

\subsection{Absolute calibration using vertical laser shots}

The drum technique for absolute calibration has been checked
for some pixels using remote laser shots at 337 and 355
nm~\cite{Roberts:2003yi,Knapik:2007yd}.
A laser pulse is shot vertically
into the air with a known intensity.  
A calculable fraction of photons 
is scattered to the aperture of the FD detector. This
yields a known number of photons arriving to the detector
for each pixel, which views a segment of the laser beam.  The
response of each pixel to the known number of photons
constitutes an absolute end-to-end calibration for those
pixels.  A calibrated laser probe is used to measure the
number of photons in typical laser pulses in the field
calibration.  A nitrogen laser (337~nm) at a distance of 4~km 
is well suited for such purposes, because the scattered light flux is in the
correct range and uncertainties due to aerosols are
minimal. The scattering angles are just greater than $90^{\circ}$, 
where the differential scattering cross section is
minimized for aerosols.  Moreover, at a distance of 4~km,
the extra scattering from the beam by aerosols is
approximately canceled by the extra aerosol attenuation of
the primarily Rayleigh-scattered light flux.
Therefore, the roving laser provides an independent and redundant absolute
calibration of the FD cameras, based on the measured light
injected into the atmosphere. 
Using the recorded laser-induced ADC traces
and the drum calibration constants, we can then compare the
number of predicted to measured photons. 

\begin{figure}[th]
\begin{center} 
\includegraphics[width=0.8\textwidth]{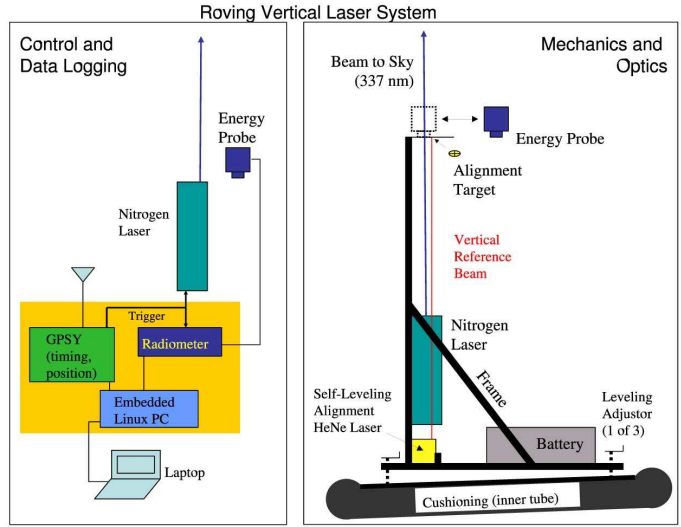}
\end{center}
\caption{Roving vertical nitrogen laser hardware and control
system.} 
\label{rovinglaser}
\end{figure}

Schematics of the hardware and controlling electronics for the
roving laser system are shown in Fig.~\ref{rovinglaser}. The
roving laser system presently in use is based on a 337~nm
nitrogen laser, providing about 100~$\mu$J per pulse, while the output 
is inherently unpolarized. The RMS
values of 100 shots in a set, fired over 200 seconds, is
typically 3\%.  Variations  of average energy in a 100 shot group
is ~2\% on the time scale of 15 minutes, and 10\% from night to
night, as measured with the energy probe inserted into the beam
for 100 shots before and after the calibration shots.  
A self-leveling He-Ne laser
provides a reference beam within $0.03^\circ$ of  vertical,
allowing alignment of the nitrogen laser to within $0.1^\circ$ of
vertical.  

An embedded Linux PC, communicating via serial connection to a
programmable GPS timing module, controls the laser firing time
relative to the GPS second.  The time stamp is used to identify
calibration events for  reconstruction. 


For comparison with the 375~nm drum calibration, the measured
relative FD response at the two calibration  wavelengths must be
known.
The relative response has been measured, and is described in the next
section.  

Overall uncertainties of the roving laser calibration 
have been assessed at 12\,\%, dominated by
laser probe calibration and atmospheric effects.  

\subsection{Multi-wavelength calibration}

For calibration at wavelengths spanning the FD acceptance, a
xenon flasher is mounted at the back of the drum, with a filter
wheel containing 5 notch filters for selection of wavelengths. 
The xenon flasher \cite{Perkins-Elmer} provides 0.4~mJ optical
output per pulse covering a broad UV spectrum, in a time period
of a few hundred nanoseconds.  A focusing lens at the filter
wheel output maximizes the intensity through the filter wheel and
into the light pipe.

\begin{figure}
\begin{center}
\includegraphics [width=0.8\textwidth]{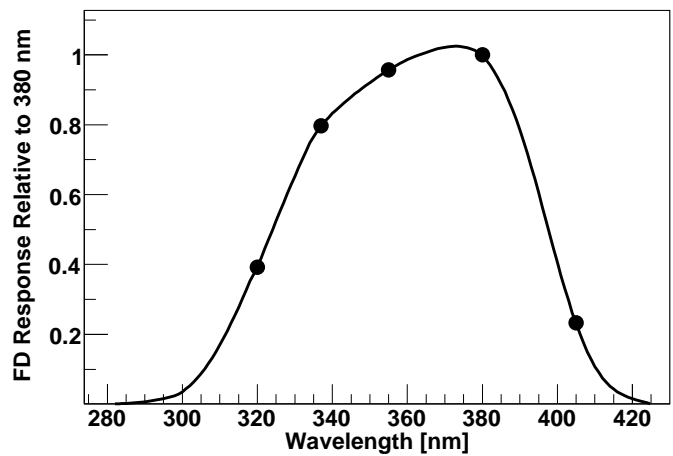}
\end{center}
\caption{
The results of the multi-wavelength measurements (see text,
and \cite{multiwavelength}.)
}
\label{multi}
\end{figure}

Relative drum intensity measurements at wavelengths of 320, 337, 355, 
380
and 405 nm have been made with the same reference PMT used in the 
absolute
measurements.  At each wavelength the recorded response from the 
reference
PMT, combined with the PMT quantum efficiency and corrected for light
source and filter width effects, yield a quantity that is proportional 
to
the number of photons emitted from the drum.  The FD response detected
using the various filters, with the drum placed in the aperture, can be
combined with the results from the laboratory to form the curve of
relative camera response \cite{multiwavelength} shown in figure \ref{multi}.

The curve in figure \ref{multi} is the result of an iterated spline fit beginning
with a response curve predicted from manufacturer's specifications for 
FD
components.  The shape of this initial curve is dominated by the FD PMT 
QE
and the UV filter transmission.  The final fit includes effects of the
notch filter transmission widths (15~nm FWHM), the reference PMT QE, 
the
xenon light source emission spectrum, and the relative drum intensity 
for
each filter, all measured in the lab, and the observed FD response to 
the
drum for each filter.  The relative uncertainty at each wavelength on 
the
curve is 5\%.


\subsection{FD relative calibration system}

The relative optical calibration system~\cite{Knapik:2007yd} is
used to monitor detector response and to track absolute
calibration between drum calibrations. The system is used
before and after each night of operation. Three positions (A, B,
and C) are illuminated for each camera, monitoring different
groups of detector components. Light is distributed through
optical fibers, from permanently installed light sources.  All
components use quartz optics.

\begin{figure}[h!]
\begin{center}
\includegraphics[width=0.8\textwidth]{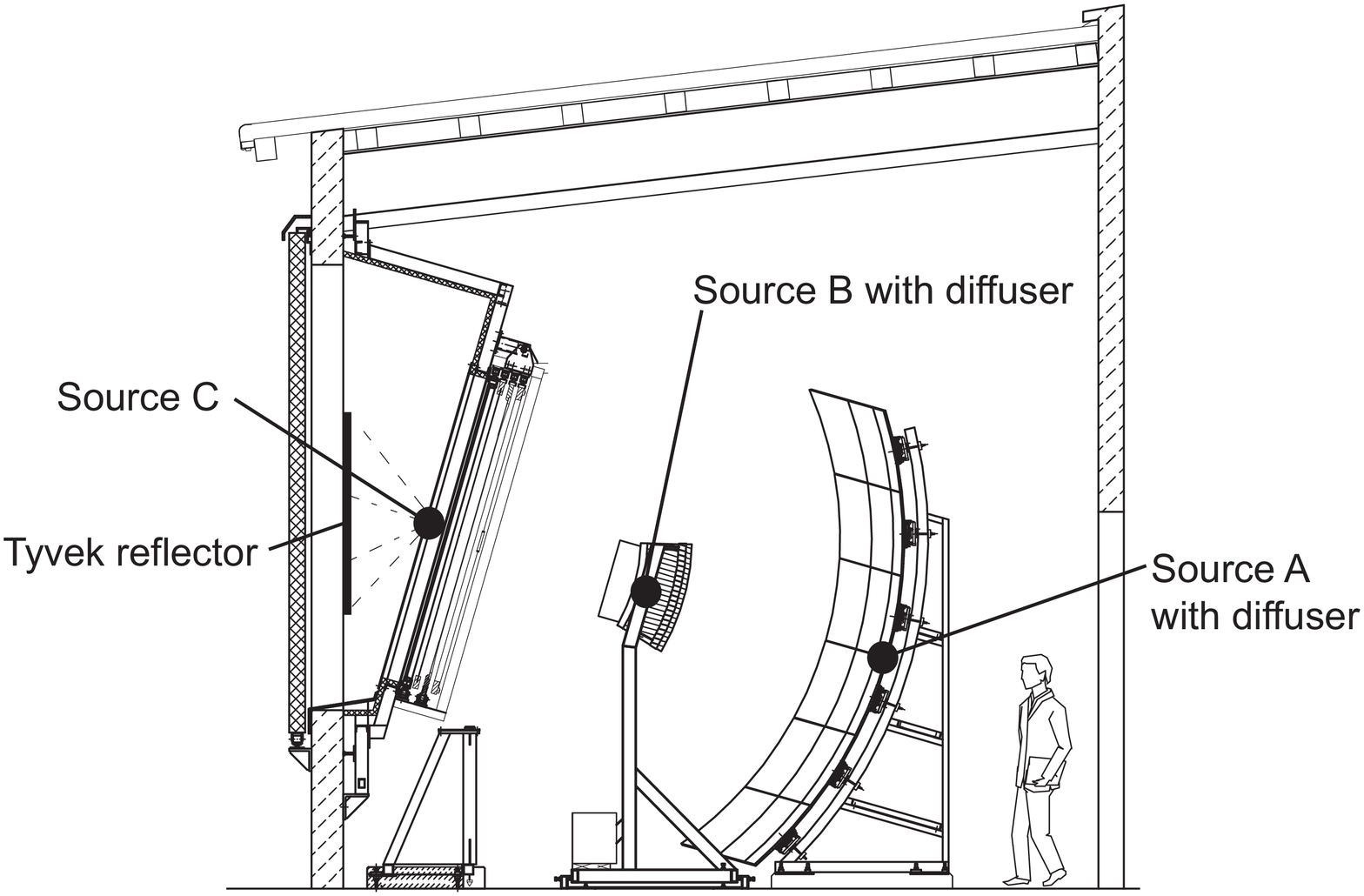}
\caption{A schematic showing positions of light sources for three different 
relative calibrations of the telescope.}
\label{relcal}
\end{center}
\end{figure}

The A fiber light source is a 470~nm LED~\cite{luxeon}.  A 7:1
splitter at the source provides light to 6 fibers, running to a 1
mm thick Teflon diffusor located at the center of each mirror in
the FD building, illuminating the camera face.  The remaining
fiber provides light for an output monitoring photodiode at the
source.  The A fiber LED is driven by a constant current source
circuit.  Normal operation pulses the LED for a series of 60~$\mu$s pulses.  

Each of the B and C light sources are xenon flash
lamps~\cite{Perkins-Elmer}. Each source is mounted at the focus
of a f/1.5 lens, with downstream optics including a beam splitter
(for a source monitoring fiber), a filter wheel, and an f/2.4
lens focusing onto a 7:1 splitter.  The 7 fibers run to each
fluorescence telescope, and to a monitor for each output, as for
the A fiber.  The B source fibers are split near each camera and
terminate at 1~mm thick Teflon diffusors located at the sides of
the camera, with the light directed at the mirror. The C source
fibers are also split, and terminate outside the aperture with
the light directed outwards.  Tyvek sheets are mounted on the
inside of the aperture shutters.  The sheets are positioned such
that they are opposite the fiber ends when the shutters are
closed, and the diffuse light scattered off the Tyvek enters the
aperture. 

The B source includes a Johnson-U filter, approximating the full
wavelength acceptance of the fluorescence telescopes.  The C
source filter wheel containing interference filters is centered at
wavelengths of 330, 350, 370, 390 and 410~nm, for monitoring
detector stability at wavelengths spanning the spectral
acceptance.

\section{Performance, operation and monitoring of the detector}\label{performance}


All four fluorescence sites have been completed and are in
operation. Los Leones has been in full operation since March 2004
and Coihueco since July 2004. Los Morados began data acquisition
in April 2005, and the fourth site at Loma Amarilla started its
operation in February 2007. 

\subsection{Uptime Fraction}

The operation of the fluorescence detector can be characterized
by the uptime, or the fraction of the total time in which the FD
was acquiring data.
Anything that disables the measurement is considered as dead
time. The main contributions to the dead time are the presence of the sun 
and nearly full moon on the sky, poor weather and the presence of the moon in
any phase within $5^\circ$ in the FOV of a telescope. The position
of the moon can be calculated in advance and the shutters of
individual telescopes are closed when the moon approaches.

The shutters are also automatically closed when the weather
conditions become dangerous for operation (high wind speed, rain,
snow, etc.) and when the observed sky brightness (caused mainly
by scattered moonlight) is too high. The influence of weather
effects depends on the season, with the worst conditions
typically occurring during Argentinian summer. Other periods of
dead time are caused by the activity of atmosphere monitoring
instruments, mostly by lidar stations and the Central Laser
Facility, readout of the electronics, and any hardware or
software problems.

\begin{figure}[tbh]
\begin{center}
\includegraphics[width=0.8\textwidth]
{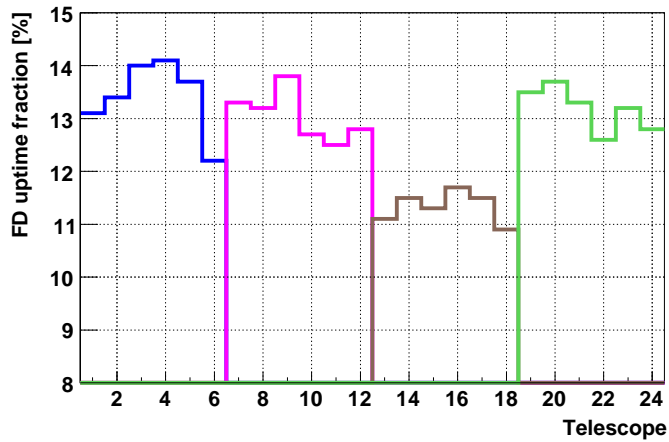}
\caption{
Uptime fraction for fluorescence telescopes. Telescopes 
are numbered as follows: Los Leones (south) site 1-6, Los Morados (east) site 7-12, Loma Amarilla 
(north) site 13-18, Coihueco (west) site 19-24.}
\label{Uptime}
\end{center}
\end{figure}

The value of uptime fraction has been derived from the data and
is cross-checked by several techniques. The average uptime
fraction for the whole observation period is around 13\,\% of the
total time since operations commenced. 
Averaged uptime fractions for individual telescopes are shwon in Fig.~\ref{Uptime}. 
For Los Leones and Coihueco the values refer to the period January 
2005 to January 2008, for Los Morados and Loma Amarilla the average 
is calculated from the individual start of operation till January 2008.

The Loma Amarilla building does not yet
have  a dedicated power line. The site is powered by a generator,
which is less reliable and has caused a lower uptime compared to
other FD sites. 

\subsection{Background conditions}

The presence of the moon above the horizon increases the
background light level, which has to be monitored. The direct
current induced by the background is eliminated by the AC
coupling of the PMT base, but it is possible to determine
background levels using
the direct relation between the fluctuations of the sky
background and the photon flux. The analysis of the fluctuations
in the ADC signal (variance analysis) performed for each night of
data taking is used to monitor the FD background signal and data
taking conditions. Thus the amount of light is derived from ADC
variance value and it is given in units (ADC counts)$^2$.

The total background signal is the sum of the electronics
background (photomultiplier and electronics noise) and the sky
brightness (airglow, moonlight, stars and planet light, zodiacal light, twilight
and artificial light) \cite{IEEE02}. The typical values
of background signals are:
3-5 (ADC counts)$^2$ for electronics background, around 20 (ADC
counts)$^2$ for cloudy nights, and between 25 and 60 (ADC
counts)$^2$ for clear moonless nights. During nights when the
moon is above the horizon, the light background level can reach
several hundred (ADC counts)$^2$. 

The optimal background conditions for observation range from 25 to 60 (ADC
counts)$^2$ which corresponds to  photon background flux from
approximately 100 to 250 photons m$^{-2}$deg$^{-2}$$\mu$s$^{-1}$.
Under such conditions, one event per 2 hours is recorded for which the
energy is determined to within 20\,\% and the depth of maximum measured
with an uncertainty of better than  40~g/cm$^2$. These high-quality events
are the ones used for physics analysis.

\subsection{Standard Operation}

The FD operation is not fully automated and at present the
assistance of a shift crew of at least two people per night is
necessary. Their responsibilities consist of several activities
before, during and after measurement each night. These include
relative calibration of the cameras and optical components before
and after observation, starting and stopping data taking
according to weather conditions, prompt correction of software or
hardware defects, etc. The operation of the FD is still evolving,
and the software development is steadily transferring the
responsibility of the human crew to automatic operation. The
ultimate aim is to operate FD telescopes in fully automatic and
remotely controlled mode from abroad.

FD is operated in nights with moon fraction below 60\,\%
beginning at the end of the astronomical twilight till the
beginning of the next astronomical twilight.
These criteria have evolved over time; before January
2005 the maximal illuminated moon fraction was 50\%, and hence
data  taking period was 2 days shorter per month. The observation
period lasts 16 days per month, with an average observational
time of about 10 hours (ranging from about 14 hours in June to 5
hours in December). 

\subsection{Slow Control System}

%

The fluorescence detectors are operated from the central campus in
Malarg\"{u}e, and are not operated directly from the FD
buildings. Therefore, the main task of the Slow Control System
(SCS) is to ensure a secure remote operation of the FD system.
The SCS works autonomously and continuously monitors detector and
weather conditions.
Commands from the remote operator are accepted only if they do
not violate safety rules that depend on the actual experimental
conditions: high-voltage, wind speed, rain, light levels
inside/outside the buildings, etc. In case of external problems
such as power failures or  communication breakdowns the SCS
performs an orderly shutdown, and also a subsequent startup of
the fluorescence detector system if the conditions have changed.
If parts of the SCS itself fail, the system automatically reverts
to a secure mode as all potentially critical system states (open
shutters, high-voltage on, etc.) have to be actively maintained. 

\begin{figure}[tb]
\begin{center}
\includegraphics[width=0.7\textwidth]{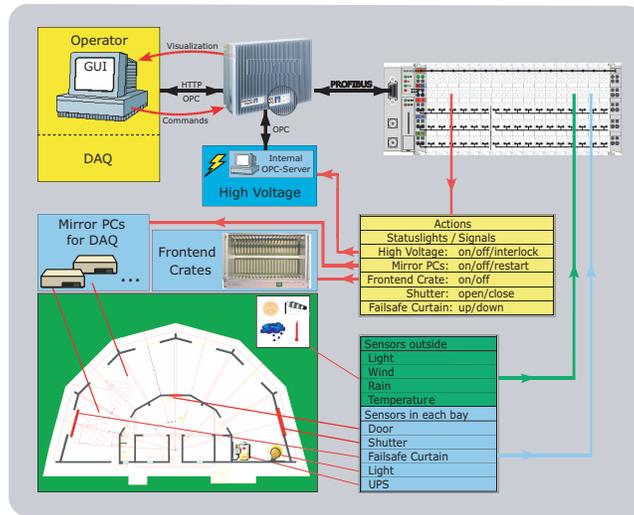} 
\caption{Schematic view of the slow-control system}
\label{scs_diagram}
\end{center}
\end{figure}

To ensure reliable supervision and to allow for high flexibility
and stability, the SCS is based on industrial PROFIBUS
components. This bus system consists of several bus terminals
with specific functions, such as analog input, digital output,
relays, etc. The terminals make up a modular bus-system which is
addressed and controlled from a PC. The slow control PC is the
central instance of the SCS in each FD building and runs the main
control software under a Windows operating system. The system for
one of the fluorescence detector buildings is sketched in Fig.
\ref{scs_diagram}. 


\begin{figure}[tb]
\begin{center}
\includegraphics[width=0.7\textwidth]{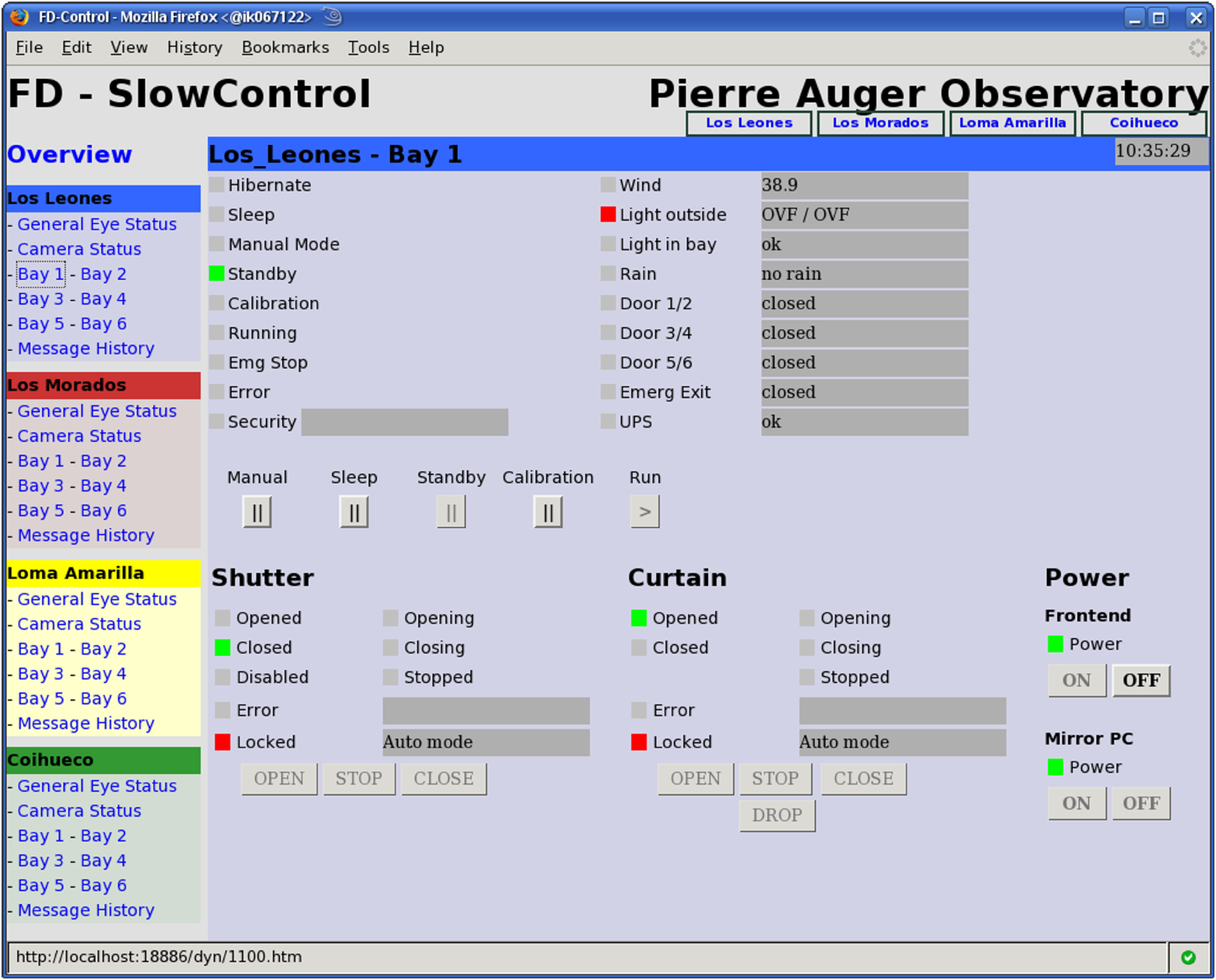}
\caption{SCS display of a FD telescope}
\label{scs_bay}
\end{center}
\end{figure}

The shift crew interacts with the control-system via
web-browsers. One central webserver communicates via OPC-XML
gateways with  4Control OPC-servers on the control-PCs. Several
views with different levels of detail display the status, from an
overview of the whole fluorescence detector of the Observatory 
down to single telescopes
(see Fig. \ref{scs_bay}). 

For the further automation of the data taking, an interface
between the Slow Control  and the Data Acquisition systems is
under development. The final goal is an automatic, scheduled
operation where the shifters have to react only to malfunctions
of the system. 

\subsection{Monitoring}
\label{sec:monitoring}


FD data-taking can only take place under specific environmental
conditions. The light-sensitive cameras must be operated on dark
nights with low wind and without rain. This makes the operation a
full task for the shifters, who judge the suitability of
operations on the basis of the information given by the SCS. The
telescope performance must be monitored constantly to assure the
quality of the recorded data, as well as guarantee the safe
operation of all detector components. A user-friendly monitoring
tool has been developed to support the shifters in judging and
supervising the status of the detector components, the
electronics, and the data-acquisition. 

\begin{figure}[b]\begin{center}%
\includegraphics[width=0.7\textwidth]{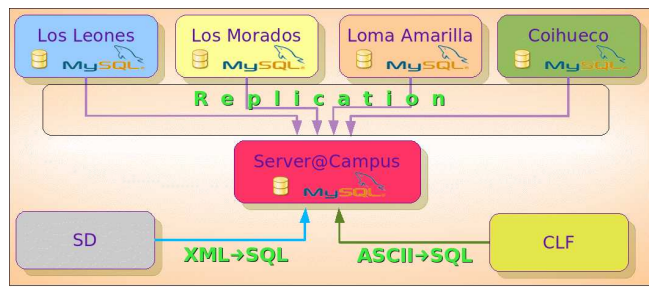}
\caption{\label {db-setup} Organisation of the databases: 
   The single databases at each FD-building are replicated to the 
   database server at the central campus, while other sources like 
   the SD insert data directly into the central database.}
\end{center}\end{figure}

\begin{figure}[bt]\begin{center}
\includegraphics[width=0.7\textwidth]{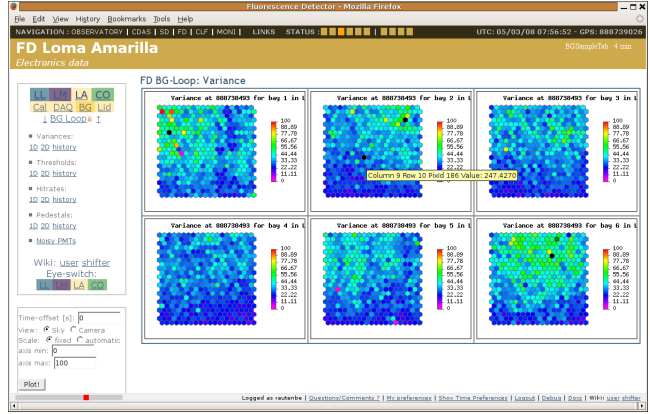}
\caption{\label {monBG2D} Screenshot of the FD monitoring web interface, showing a selection of FD background data for all
pixels in the six cameras at Loma Amarilla.} 
\end{center}
\end{figure}

The monitoring tool \cite{julian} is built upon online MySQL databases,
organized by FD building. The databases record data from regular
FD measurements and from calibrations and atmospheric surveys.
The data are  transferred to  a central server at the
Malarg\"{u}e campus using  the internal MySQL database
replication mechanism. This mechanism recognizes communication
problems and tries to synchronize database changes when the
connection is reestablished; this guarantees the completeness of
the dataset on the central server, even if the information are
not available online immediately due to network failures.
Fig.~\ref{db-setup} shows the schematic layout of the databases.

The user interface is based on a webserver running Apache. The
website uses  PHP, CSS and JavaScript; graphs and custom
visualizations are implemented using the JPGraph package for
direct PHP calls to the database. They are dynamically generated
to be accessible not only for the shifter, but also remotely for
experts. Graph explanations as well as troubleshooting tips are
available through linked wiki pages. Alarms are triggered
automatically in case of special occurrences. This tool helps the
shifter to monitor constantly the performance of the detector to
assure the quality of the recorded data as well as guaranteeing
the safe operation of all detector components. In addition it
offers a unique opportunity to monitor the long term stability of
some key quantities and the data quality. 

\subsection {Online events}

\begin{figure}[tb]
\begin{center}
\includegraphics[width=0.9\textwidth]{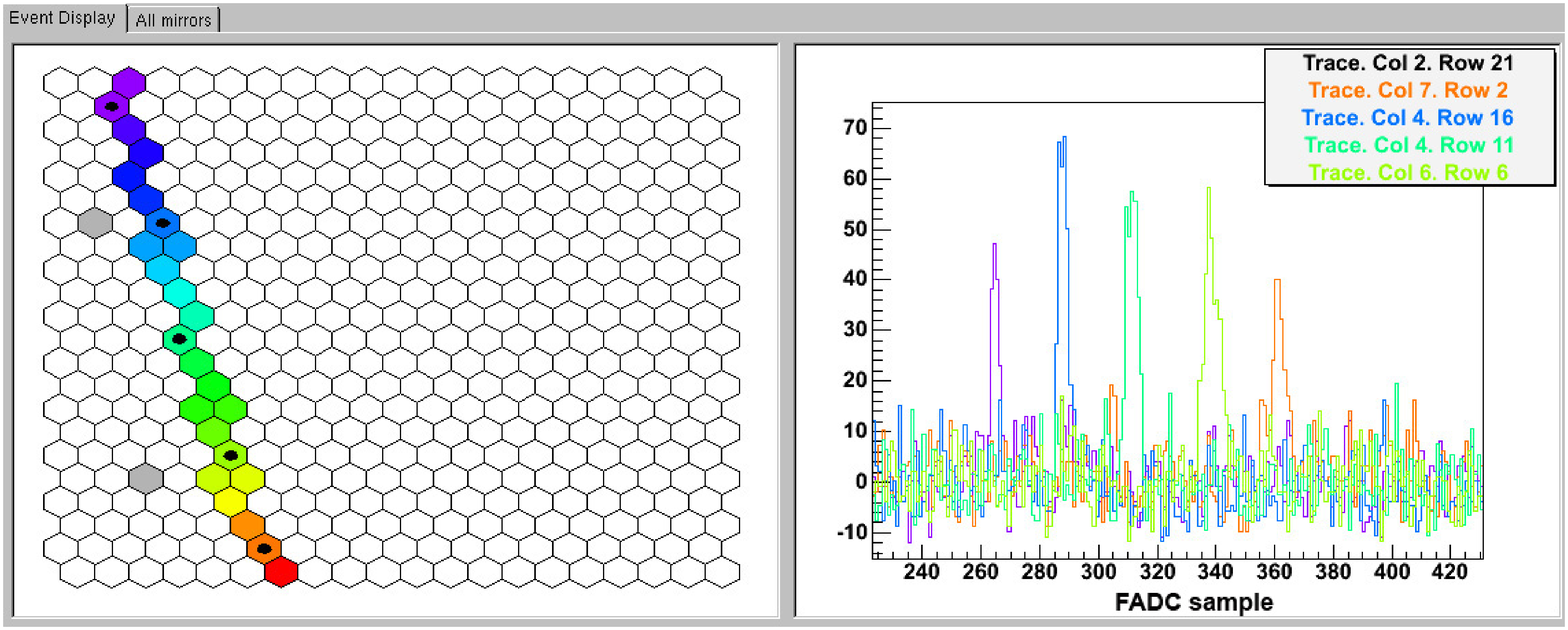}
\includegraphics[width=0.9\textwidth]{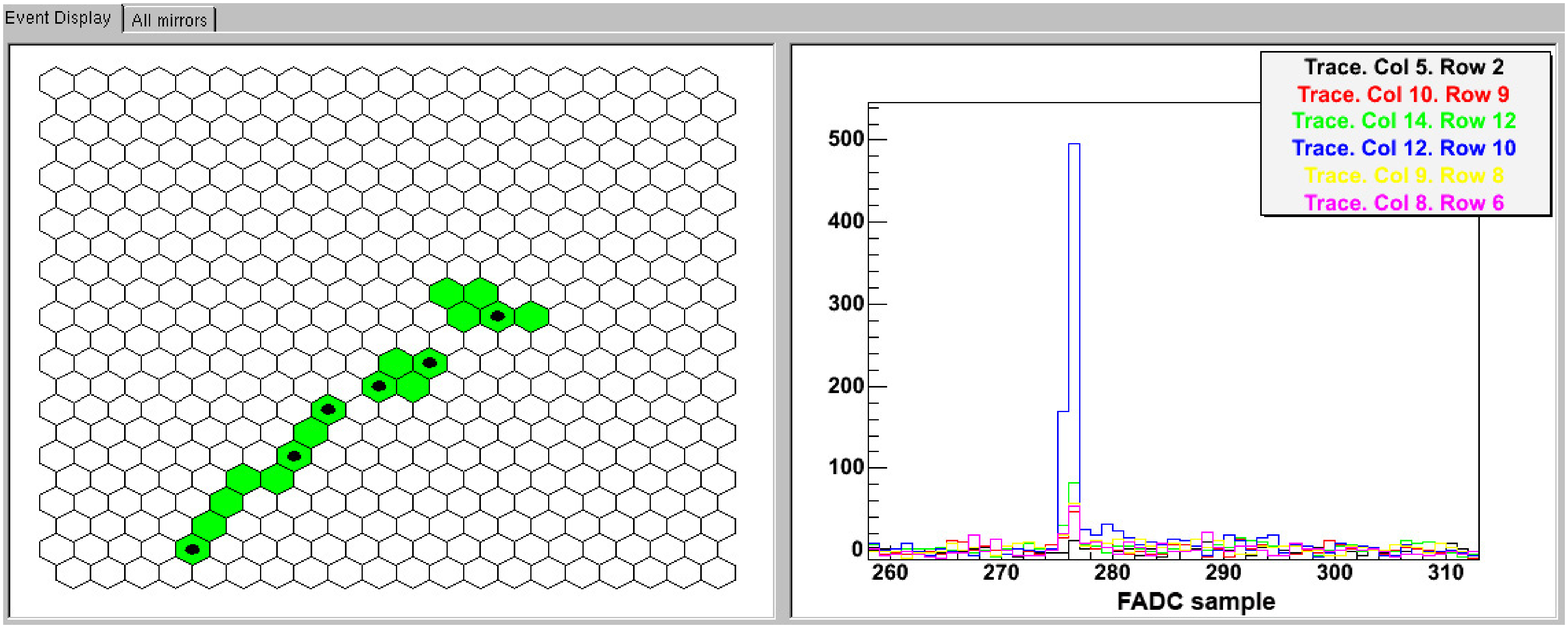}
\caption{Top: A cosmic ray shower event as it appears in the event display. The pattern of the activated pixels is shown in the left panel while the right panel exhibits the response of the selected 
pixels as a function of time for the pixels marked by a black dot. The bin size is 100 ns. The development of the shower in the atmosphere can be qualitatively seen.
Bottom: A background event most likely due to cosmic ray muon interacting with the glass of the PMT.  All activated pixels give signals at the same time, a feature which is not compatible with a cosmic ray shower.}
\label{figcamera6}
\end{center}
\end{figure}

Besides using the online monitoring system for a continuous performance 
check by integrating over the most recent history of data, the operators 
are furnished with an online event display of the cameras. Two typical 
event examples corresponding to a real cosmic ray air shower and to 
a background event are shown in Fig.~\ref{figcamera6}. 
The light spot of the air  shower, shown in the upper part of the figure, 
takes about 10~$\mu$s to cross the FOV of the camera while individual pixels 
are illuminated for up to about 1~$\mu$s. This is clearly different from the 
event shown in the bottom part of the figure. Here, all activated pixels show 
huge narrow pulses all starting in the same sample of the 10 MHz 
ADC. Such a geometrical pixel pattern and timing feature is 
incompatible with an air shower and is most likely due to a muon penetrating 
the camera.

\section{Event Reconstruction}\label{reconstruction}

\subsection{Geometrical Reconstruction}


A hybrid detector achieves the best geometrical accuracy by using
timing information from all the detector components, both FD
pixels and SD stations.
Each element records a pulse of light from which it is possible to determine 
the time of the pulse and its uncertainty. Each trial geometry for the shower 
axis yields a prediction for the signal arrival times at each detector component.

%
Differences between actual and predicted
times are weighted using their corresponding uncertainties,
squared, and summed to construct a $\chi^{2}$ value. The
hypothesis with the minimum value of $\chi^{2}$ is the
reconstructed shower axis.

\begin{figure}[!ht]
\begin{center}
\includegraphics[width=0.6\textwidth]{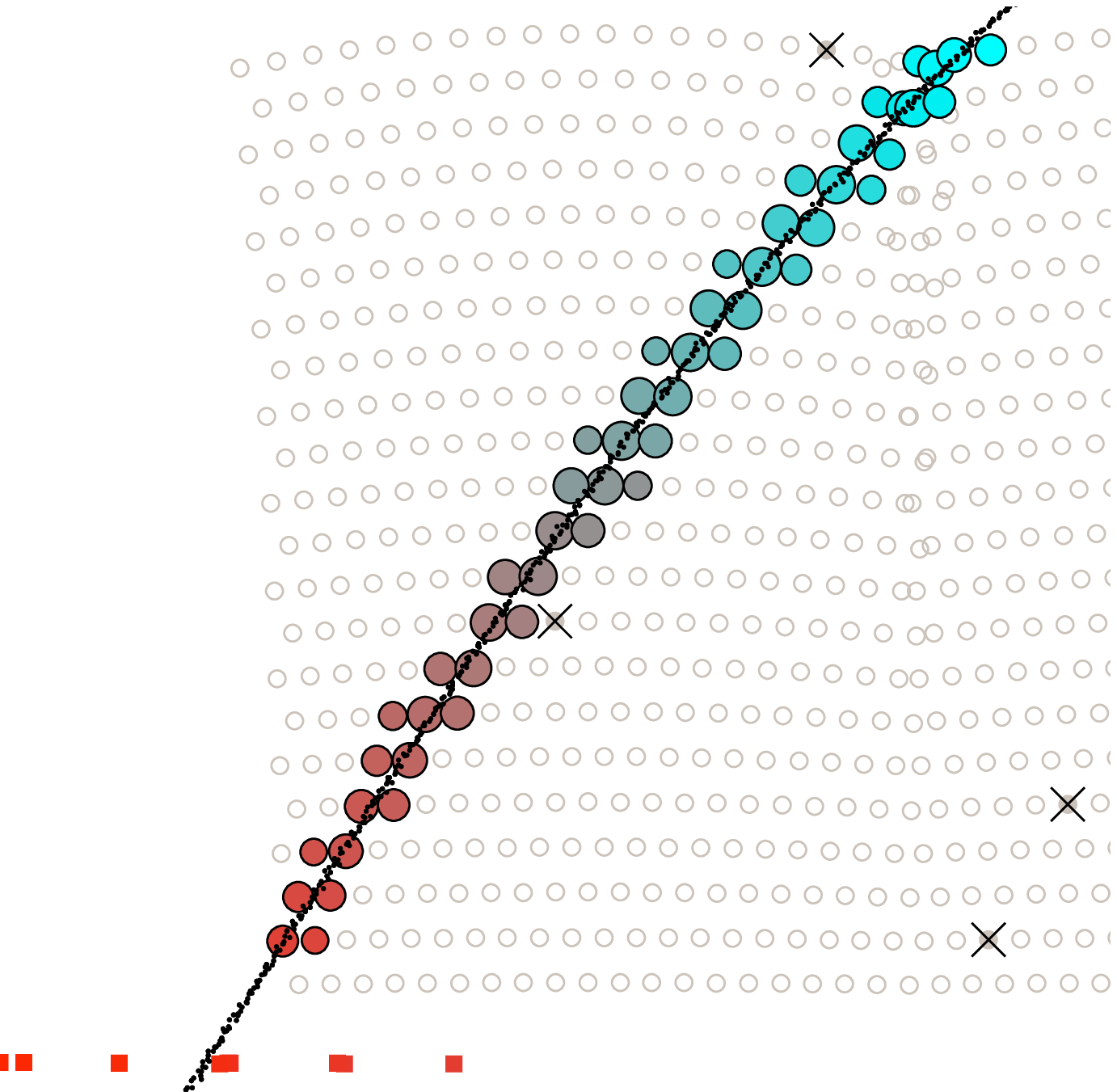}
\caption{Light track of a hybrid event as seen by the fluorescence telescopes.
The different colors indicate the timing sequence of the triggered pixels.
The full line is the fitted shower-detector plane. (See text for explanation.)
The red squares in the bottom left represent the surface stations that also triggered in this event.
The crosses mark camera pixels that had a signal within the time of the trigger, but were marked by the reconstruction algorithm as too far either in distance (to the shower-detector plane) or in time (to the time fit).}
\label{fig:Gcamera} 
\end{center}
\end{figure}

\begin{figure}[!ht]
\begin{center}
\includegraphics[width=0.6\textwidth]{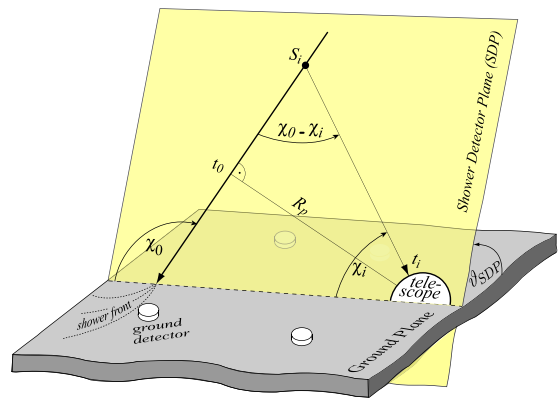}
\caption{Illustration of the geometrical shower reconstruction from the observables of the fluorescence detector \cite{Kuempel:2008ba}.}
\label{fig:sketch} 
\end{center}
\end{figure}

In the FD, cosmic ray showers are detected as a sequence of
triggered pixels in the camera. An example of an event
propagating through two adjacent FD telescopes is presented in
Fig.~\ref{fig:Gcamera}. The first step in the analysis is the
determination of the shower-detector plane (SDP). The SDP is the
plane that includes the location of the eye and the line of the
shower axis. (See the sketch in Fig.~\ref{fig:sketch}.)
Experimentally, it is the plane through the eye which most nearly
contains the pointing directions of the FD pixels centered on the
shower axis. (See fitted line in Fig.~\ref{fig:Gcamera}.) Using a
known axis provided from the Central Laser Facility (CLF),
described in Ref.~\cite{CLF}, the SDP reconstruction error can be
evaluated by comparing the space angle between the normal vector
to the experimentally determined SDP and the known true normal
vector. This uncertainty in the SDP 
is of the order of a few tenths of a degree depending on, for example, 
the length of the observed track in the camera.

Next, the timing information of the pixels is used for
reconstructing the shower axis within the SDP. As illustrated in
Fig.~\ref{fig:sketch}, the shower axis can be characterized by
two parameters: the perpendicular distance $R_{p}$ from the
camera to the track, and the angle $\chi_0$ that the track makes
with the horizontal line in the SDP. Each pixel which observes
the track has a pointing direction which makes an angle
$\chi_{i}$ with the horizontal line. Let $t_{0}$ be the time when
the shower front on the axis passes the point of closest approach
$R_{p}$ to the camera. The light arrives at the $i^{\rm th}$ pixel at the time 

\begin{equation}
	t_i = t_0 + \frac{R_p}{c} \tan{[(\chi_0 - \chi_i)/2]}.
	\label{eq:timefit}
\end{equation}

\begin{figure}[!ht]
\begin{center}
\includegraphics[width=0.9\textwidth]{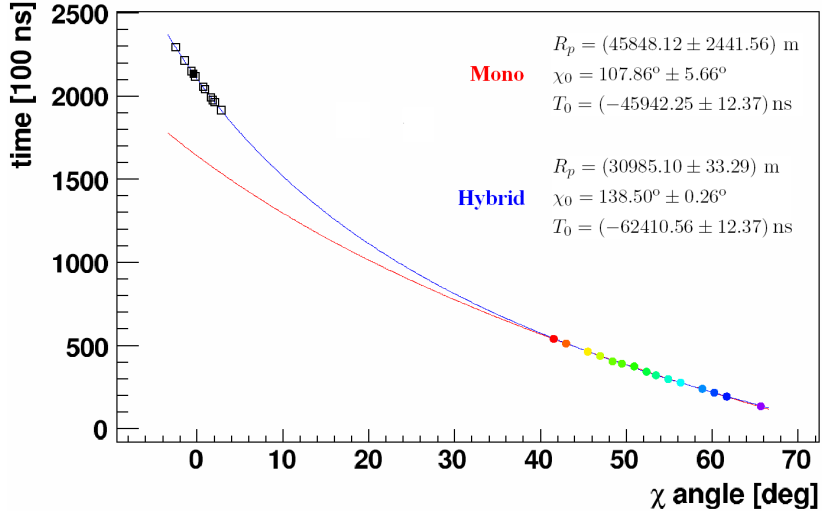}
\caption{Functional form that correlates the time of arrival of the light at each pixel with the 
angle between the pointing direction of that particular pixel and the horizontal line within the shower-detector plane.
FD data (color points) and SD data (squares) are superimposed to the monocular (red line) and hybrid (blue line) reconstruction fits.
The full square indicates the SD station with the highest signal.
This is a typical event in which the monocular reconstruction does not work well.}
\label{fig:timefit} 
\end{center}
\end{figure}

The shower parameters are determined by fitting the data points
to this functional form. Using the 
fast sampling electronics, such a
monocular reconstruction may achieve excellent accuracy. However,
the accuracy of the monocular reconstruction is limited when the
measured angular speed $d\chi/dt$ does not change much over the
observed track length. An example is shown in
Fig.~\ref{fig:timefit}. For these events (usually short tracks)
there is a small curvature in the functional form of
Eq.~(\ref{eq:timefit}) such that there is a family of possible
$(R_p,\chi_0)$ axis solutions. 
$R_p$ and $\chi_0$ are tightly correlated, but neither value is well
constrained.  This leads to uncertainty in other shower
parameters, including the reconstructed shower energy.

The fit degeneracy can be broken by combining the timing
information from the SD stations with that of the FD telescopes.
This is called the \textit{hybrid} reconstruction. The hybrid
solution for the example shown in Fig.~\ref{fig:timefit} is shown
as a blue line and the uncertainties in the parameters are specified in the legend.


Since the SD operates with a $100\%$ duty cycle, most of the
events observed by the FD are in fact hybrid events. There are
also cases where the fluorescence detector, having a lower energy
threshold, promotes a sub-threshold array trigger (see
section~\ref{subsubsec:hybridT3}). Surface
stations are then matched by timing and location. This is an
important capability because these sub--threshold hybrid events
would not have triggered the array otherwise. In fact, the time
of arrival at a single station at ground can suffice for the
hybrid reconstruction. 

The reconstruction uncertainties are validated using events with
\textit{known} geometries, i.e.\ light scattered from laser pulses. Since the
location of the CLF (approximately equidistant from the first
three fluorescence sites) and the direction of the laser beam are
known to an accuracy better than the expected angular resolution
of the fluorescence detector, laser shots from the CLF can be
used to measure the accuracy of the geometrical reconstruction.
Furthermore, the laser beam is split and part of the laser light
is sent through an optical fiber to a nearby surface array
station. Thus, the axis of the laser light can be reconstructed
both in monocular mode and in the \textit{single-tank} hybrid
mode.

%
%

Using the timing information from the telescope pixels together
with the surface stations to reconstruct real air showers, a core
location resolution of $50$~m is achieved. The resolution for the
arrival direction of cosmic rays is
$0.6^{\circ}$~\cite{resolution}. These results for the hybrid
accuracy are in good agreement with estimations using analytic
arguments~\cite{Sommers:1995dm}, measurements on real data using
a bootstrap method~\cite{Fick:2003qp}, and previous simulation
studies~\cite{Dawson:1996ci}.

\subsection{Shower Profile and Energy Reconstruction}

\begin{figure}[tbh]
\begin{center}
\includegraphics[width=0.8\textwidth]{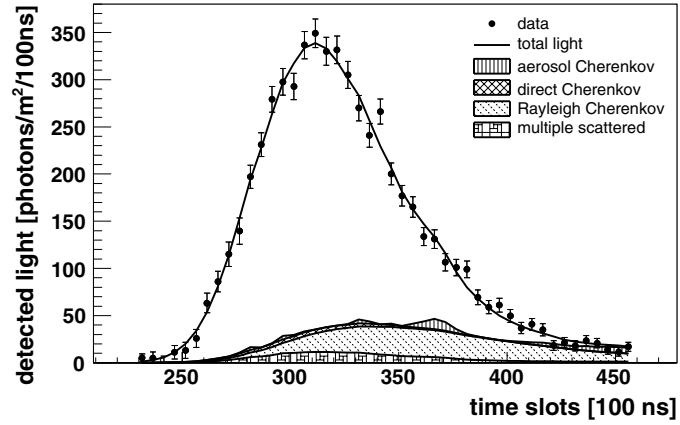}
\caption{\label{fig:lightAtAperture} Example of a light-at-aperture measurement (dots) and reconstructed light sources (hatched areas).}
\end{center}
\end{figure}
\begin{figure}[tbh]
\begin{center}
\includegraphics[width=0.8\textwidth]{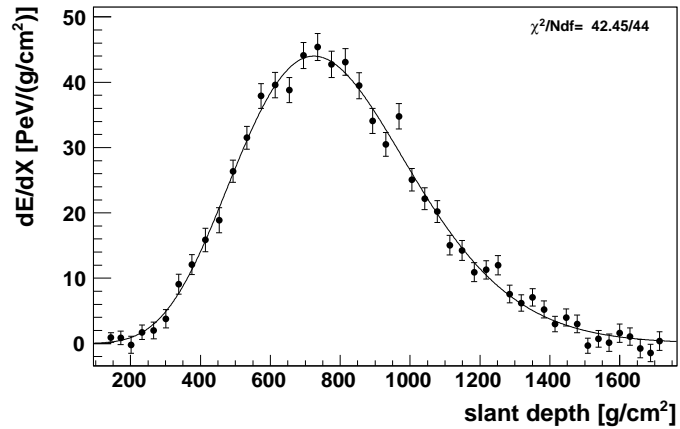}
\caption[profile]{\label{fig:dEdXProfile}Energy deposit profile
reconstructed from the light at aperture shown in
Fig.~\ref{fig:lightAtAperture}. The line shows a Gaisser-Hillas fit of the profile.  The energy reconstruction for this shower was $3.0 \pm 0.2 \cdot 10^{19}$~eV.}
\end{center}
\end{figure}

Once the geometry of the shower is known, the light collected at
the aperture as a function of time can be converted to energy
deposit at the shower as a function of slant depth. For this
purpose, the light attenuation from the shower to the telescope
needs to be estimated and all contributing light sources need to
be disentangled~\cite{profileReco}: fluorescence
light~\cite{Nagano:2004am,airflyAtm,airflydEdX,gorafluo}, direct and
scattered Cherenkov light~\cite{cher:giller,cher:nerling} as well
as multiple-scattered light~\cite{Roberts:2005xv,pekalamulti}.\\ An example
of the measured light at the telescope aperture and the
reconstructed light contributions and energy deposit profile is
shown in Figs.~\ref{fig:lightAtAperture}
and~\ref{fig:dEdXProfile}.\\ The calorimetric energy of a shower
is estimated by fitting a Gaisser-Hillas function~\cite{ghfunc}
to the reconstructed energy deposit profile and integrating it.
Finally, the total energy of the shower is obtained by correcting
for the 'invisible energy' carried away by neutrinos and high
energy muons \cite{invE:barbosa}.  After quality selection, the
energy resolution (defined as event-to-event statistical uncertainty) 
of the fluorescence detector is $\le$ 10\%~\cite{Dawson:2007di}.

\subsection{Detector Exposure}


The detection volume of fluorescence detectors varies with
energy, as showers with higher energies emit more light and can
be detected further away from the detector. The aperture also
depends on environmental factors such as night-sky background
light and atmospheric conditions. To model these
effects, large sets of detailed MC simulations are used. The
response of the Auger fluorescence telescopes has been simulated
and the detector aperture has been estimated as a function of
energy, average atmospheric conditions, and primary cosmic ray
mass for a fixed configuration: a fully built detector with four
fluorescence detectors and a 3000 km$^{2}$ surface array
\cite{prado, bellido}. The aperture increases from approximately
$900~\mathrm{km^2~sr}$ at $10^{17.5}~\mathrm{eV}$ to about
$7400~\mathrm{km^2~sr}$ at $10^{19}~\mathrm{eV}$. 

However, the more important quantity here is the {\it hybrid
detector exposure} which  accounts for both the time variability
of the atmospheric conditions and the detector setup. Very
detailed time-dependent detector Monte Carlo simulations are used
to reproduce the actual data taking conditions of all components
of the Pierre Auger Observatory and to derive the hybrid exposure.
To assure good agreement between data and Monte Carlo, extensive
comparisons are performed at all reconstruction levels.

To limit the influence of trigger threshold effects, an
energy-dependent fiducial volume has been defined and strict
quality criteria are applied during the determination of the
exposure~\cite{Perrone:2007he}. For example, only data with a
successful hybrid geometry reconstruction are selected, and the
SD station used for the reconstruction has to lie within
$750~\mathrm{m}$ of the shower axis. This condition ensures that
the probability to trigger at least one surface station is 
almost equal to one, and it significantly  reduces the influence of the
a-priori unknown primary mass composition. The resulting hybrid
exposure, after all selection cuts accumulated during the 3 years
of the building phase of the observatory, increases from about
$50~\mathrm{km^2~sr~yr}$ at $10^{18}~\mathrm{eV}$ to about
$550~\mathrm{km^2~sr~yr}$ at $10^{19.5}~\mathrm{eV}$.





\section{Summary}\label{conclusions}

The Observatory has been in scientific operation
since late 2003, and the synergy between the surface array
and the fluorescence detector has proved to be most
fruitful.  In particular, the calorimetric FD energy
measurements have provided a calibration for the
high-statistics data set obtained with full-time operation
of the surface array.

This paper has described the Auger
fluorescence detector system - its hardware, performance,
calibration, and event reconstruction methods.  The 24
wide-field telescopes view the atmosphere above the entire
3000-km$^2$ surface array.  The Auger Observatory has opened a
new chapter in cosmic ray physics because of both the large
number of recorded high energy air showers and the quality
of the hybrid measurements.  The growing data set is
expected to resolve the important questions relating to the
highest energy particles of the universe.

\section{ Acknowledgments}
The successful installation and commissioning of the Pierre Auger Observatory
would not have been possible without the strong commitment and effort
from the technical and administrative staff in Malarg\"ue.

We are very grateful to the following agencies and organizations for financial support: 
Comisi\'on Nacional de Energ\'ia At\'omica, 
Fundaci\'on Antorchas,
Gobierno De La Provincia de Mendoza, 
Municipalidad de Malarg\"ue,
NDM Holdings and Valle Las Le\~nas, in gratitude for their continuing
cooperation over land access, Argentina; 
the Australian Research Council;
Conselho Nacional de Desenvolvimento Cient\'ifico e Tecnol\'ogico (CNPq),
Financiadora de Estudos e Projetos (FINEP),
Funda\c{c}\~ao de Amparo \`a Pesquisa do Estado de Rio de Janeiro (FAPERJ),
Funda\c{c}\~ao de Amparo \`a Pesquisa do Estado de S\~ao Paulo (FAPESP),
Minist\'erio de Ci\^{e}ncia e Tecnologia (MCT), Brazil;
AVCR AV0Z10100502 and AV0Z10100522,
GAAV KJB300100801 and KJB100100904,
GACR 202/06/P006,
MSMT-CR LA08016, LC527, 1M06002 and  MSM0021620859, Czech Republic;
Centre de Calcul IN2P3/CNRS, 
Centre National de la Recherche Scientifique (CNRS),
Conseil R\'egional Ile-de-France,
D\'epartement  Physique Nucl\'eaire et Corpusculaire (PNC-IN2P3/CNRS),
D\'epartement Sciences de l'Univers (SDU-INSU/CNRS), France;
Bundesministerium f\"ur Bildung und Forschung (BMBF),
Deutsche Forschungsgemeinschaft (DFG),
Finanzministerium Baden-W\"urttemberg,
Helmholtz-Gemeinschaft Deutscher Forschungs- zentren (HGF),
Ministerium f\"ur Wissenschaft und Forschung, Nordrhein-Westfalen,
Ministerium f\"ur Wissenschaft, Forschung und Kunst, Baden-W\"urttemberg, Germany; 
Istituto Nazionale di Fisica Nucleare (INFN),
Ministero dell'Istruzione, dell'Universit\`a e della Ricerca (MIUR), Italy;
Consejo Nacional de Ciencia y Tecnolog\'ia (CONACYT), Mexico;
Ministerie van Onderwijs, Cultuur en Wetenschap,
Nederlandse Organisatie voor Wetenschappelijk Onderzoek (NWO),
Stichting voor Fundamenteel Onderzoek der Materie (FOM), Netherlands;
Ministry of Science and Higher Education,
Grant Nos. 1 P03 D 014 30, N202 090 31/0623, and PAP/218/2006, Poland;
Funda\c{c}\~ao para a Ci\^{e}ncia e a Tecnologia, Portugal;
Ministry for Higher Education, Science, and Technology,
Slovenian Research Agency, Slovenia;
Comunidad de Madrid, 
Consejer\'ia de Educaci\'on de la Comunidad de Castilla La Mancha, 
FEDER funds, 
Ministerio de Ciencia e Innovaci\'on,
Xunta de Galicia, Spain;
Science and Technology Facilities Council, United Kingdom;
Department of Energy, Contract No. DE-AC02-07CH11359,
National Science Foundation, Grant No. 0450696,
The Grainger Foundation USA; 
ALFA-EC / HELEN,
European Union 6th Framework Program,
Grant No. MEIF-CT-2005-025057, 
European Union 7th Framework Program, Grant No. PIEF-GA-2008-220240
and UNESCO.


\begin{thebibliography}{99}
\bibitem{general}
   A.~A.~Watson,
 Proc. 30th ICRC (2007), arXiv:0801.2321 [astro-ph].

\bibitem{SD}
  I.~Allekotte {\it et al.}  [Pierre Auger Collaboration],
  Nucl.\ Instrum.\ Meth.\  A {\bf 586} (2008) 409.

\bibitem{sybenzvi}
J. Abraham {\it et al.} [Pierre Auger Collaboration], submitted to Astropart. Phys (2009).

\bibitem{flyseye}
R.M. Baltrusaitus {\it et al.}, Nucl. Instrum. Meth. Phys. Res. A {\bf 240} (1985) 410.

\bibitem{hires}
T. Abu-Zayyad {\it et al.}, Nucl Instrum. Meth. Phys. Res. A {\bf 450} (2000) 253.

\bibitem{telarray}
  H.~Kawai {\it et al.},
  Nucl.\ Phys.\ Proc.\ Suppl.\  {\bf 175-176} (2008) 221.

\bibitem{nagano}
M. Nagano {\it et al.}, Astropart. Phys. {\bf 22}, 235 (2004).

\bibitem{airfly}
M. Ave {\it et al.}, Astropart. Phys. {\bf 28}, 41 (2007).

\bibitem{CLF}
B.~Fick {\it et al.}  [Pierre Auger Collaboration],
  JINST {\bf 1} (2006) P11003.


\bibitem{Schott} Schott Glaswerke, Mainz, Germany ({\tt http://www.schott.com}).




\bibitem{malacara}
D. Malacara, Optical shop testing, 3rd edition, Wiley-Interscience (2007)

\bibitem{lenses1}
 M.~A.~L.~de Oliveira, V.~de Souza, H.~C.~Reis and R.~Sato,
  Nucl.\ Instrum.\ Meth.\  A {\bf 522} (2004) 360.


\bibitem{lenses2} R.~Sato, C.~O.~Escobar [Pierre Auger Collaboration],
Proc. 29th ICRC (2005), FERMILAB-CONF-05-285-AD-E-TD.



\bibitem{Schwantz} Schwantz Ferramentas Diamantadas e Com\'ercio
\'Optico Ltda, Indaiatuba, Brazil.

\bibitem{sim1}
 S.~Agostinelli {\it et al.}  [GEANT4 Collaboration],
  Nucl.\ Instrum.\ Meth.\  A {\bf 506} (2003) 250.

\bibitem{sim2}
M.~G.~Pia  [Geant4 Collaboration],
  Nucl.\ Phys.\ Proc.\ Suppl.\  {\bf 125} (2003) 60.

\bibitem{camera}
 M.~Ambrosio {\it et al.},
  Nucl.\ Instrum.\ Meth.\  A {\bf 478} (2002) 125.

\bibitem{stars} 
C.~De~Donato {\it et al.}, Astroparticle Physics {\bf 28} (2007) 216.

\bibitem{photonis}
PHOTONIS, {\tt http://www.photonis.com}.

\bibitem{divider}
 S.~Argiro {\it et al.},
  Nucl.\ Instrum.\ Meth.\  A {\bf 461} (2001) 440.

\bibitem{testpmt} 
K.~H.~Becker {\it et al.},
Nucl.\ Instrum.\ Meth.\  A {\bf 576} (2007) 301.

\bibitem{intratec}
Intratec GmbH, Beim Haferhof 5, D-25479 Ellerau; {\tt www.intratec.de}.

\bibitem{gemmeke}

 H.~Gemmeke, A.~Grindler, H.~Keim, M.~Kleifges, N.~Kunka, D.~Chernyakhovsky and Z.~Szadkowski,
  IEEE Trans.\ Nucl.\ Sci.\  {\bf 47} (2000) 371.


\bibitem{firewire}
The FireWire interface is specified as IEEE 1394, see e.g. the book
Don Anderson, 'FireWire System Architecture, Second Edition IEEE 1394a
MindShare, Inc.', Addison Wesley, ISBN 0-201-48534-4.

\bibitem{vivi}
V. Scherini [Pierre Auger Collaboration], Proc. of the 20th European Cosmic Ray 
Symposium, Lisboa, Portugal (2006), ecrs-06-s0-187 

\bibitem{ICRC01}
H.~Gemmeke, M.~Kleifges, A.~Kopmann, N.~Kunka, A.~Menshikov, and D.~Tcherniakhovski,
Proc. 27th ICRC (2001), p737.

\bibitem{IEEE02}
M. Kleifges {\it et al.},
  IEEE TNS Vol. 50 {\bf No 4} (2003) P1204-1207.

\bibitem{szadkowski}
Z. Szadkowski,  Nucl.\ Instrum.\ Meth.\  A {\bf 465} (2001) 540.


\bibitem{NewTLT}
A. Schmidt {\it et al.},
``Third Level Trigger for the Fluorescence Telescopes of the Pierre Auger Observatory'',  subm.\ for publication to Nucl.\ Instr.\ Meth.\ A.
 
\bibitem{abscal}
  J.~T.~Brack, R.~Meyhandan, G.~J.~Hofman and J.~Matthews,
  Astropart.\ Phys.\  {\bf 20} (2004) 653.

\bibitem{Nichia} Nichia America Corp., NSHU550 UV LED.

\bibitem{luxeon}
Luxeon V Star, Document DS30, available at {\tt www.lumileds.com}.

\bibitem{NIST} National Institute of Standards and Technology,
   U.S. Dept. of Commerce,  Calibration Program, Gaithersburg, MD
   20899-2330;  NIST Special Publication 250-41, 1998.







\bibitem{Roberts:2003yi}
  M.~D.~Roberts  [Auger Collaboration],
  Proc. 28th (ICRC 2003), arXiv:astro-ph/0308410.

\bibitem{Knapik:2007yd}
  R.~Knapik {\it et al.},
Proc. 30th ICRC (2007)
  arXiv:0708.1924 [astro-ph].


\bibitem{Perkins-Elmer} 
LS-1130-4 1100 Series FlashPac with FX-1160 xenon
flash-lamp with reflector and borosilicate window from Perkin Elmer
Opto-electronics, 35 Congress St., Salem, MA 01970.

\bibitem{multiwavelength} A. Rovero {\it et al.},
for the Pierre Auger Collaboration, Astropart. Phys. {\bf 31}, 305 (2009).

\bibitem{julian}
Julian Rautenberg {\it et al.} [Pierre Auger Collaboration],
Proc. 30th ICRC (2007), Vol. {5}, p. 993





\bibitem{prado}
  L.~Prado {\it et al.},
  Nucl.\ Instrum.\ Meth.\  A {\bf 545} (2005) 632.

\bibitem{bellido}
  J.~A.~Bellido {\it et al.}  [Pierre Auger Collaboration],
  Proc. 29th ICRC (2005), arXiv:astro-ph/0507103.

\bibitem{Perrone:2007he}
  L.~Perrone  [Pierre Auger Collaboration],
  Proc. 30th ICRC (2007), arXiv:0706.2643 [astro-ph].

\bibitem{resolution}
  C.~Bonifazi  [Pierre Auger Collaboration],
Proc. 29th ICRC (2005), FERMILAB-CONF-05-301-E-TD.


\bibitem{Kuempel:2008ba}
D.~Kuempel, K.~H.~Kampert and M.~Risse,
Astropart.\ Phys.\  {\bf 30} (2008) 167
arXiv:0806.4523 [astro-ph].




\bibitem{Sommers:1995dm}
  P.~Sommers,
  Astropart.\ Phys.\  {\bf 3}, 349 (1995).

\bibitem{Fick:2003qp}
  B.~Fick  [Pierre Auger Collaboration],
  Proc. 28th ICRC (2003), arXiv:astro-ph/0308512.

\bibitem{Dawson:1996ci}
  B.~R.~Dawson, H.~Y.~Dai, P.~Sommers and S.~Yoshida,
  Astropart.\ Phys.\  {\bf 5}, 239 (1996).





 
\bibitem{profileReco} 
  M.~Unger, B.~R.~Dawson, R.~Engel, F.~Schussler and R.~Ulrich,
  Nucl.\ Instrum.\ Meth.\  A {\bf 588} (2008) 433.


\bibitem{Nagano:2004am}
  M.~Nagano, K.~Kobayakawa, N.~Sakaki and K.~Ando,
  Astropart.\ Phys.\  {\bf 22} (2004) 235.

\bibitem{airflyAtm}
M.~Ave {\it et al.} [AIRFLY Collaboration],  Nucl.\ Instrum.\ Meth.\
A {\bf 597} (2008) 50. 

\bibitem{airflydEdX}
M.~Ave {\it et al.} [AIRFLY Collaboration],  Nucl.\ Instrum.\ Meth.\
A {\bf 597} (2008) 46. 

\bibitem{gorafluo}
D~.Gora {\it et al.}, Astropart.\ Phys.\ {\bf 24} (2006) 484.

\bibitem{cher:giller} 
 M.~Giller, G.~Wieczorek, A.~Kacperczyk, H.~Stojek and W.~Tkaczyk,
  J.\ Phys.\ G {\bf 30} (2004) 97.

\bibitem{cher:nerling}
  F.~Nerling, J.~Bluemer, R.~Engel and M.~Risse,
  Astropart.\ Phys.\  {\bf 24} (2006) 421
  [arXiv:astro-ph/0506729].


\bibitem{Roberts:2005xv} 
  M.~D.~Roberts,
  J.\ Phys.\ G {\bf 31} (2005) 1291.

\bibitem{pekalamulti}
J. Pekala {\it et al.}, Nucl. Instr. Meth. A, in print (2009), arXiv:0904.3230 [astro-ph]

\bibitem{ghfunc}
T.~K. Gaisser, A.~M. Hillas, Proc. 15th ICRC (1977).

\bibitem{invE:barbosa}
  H.~M.~J.~Barbosa, F.~Catalani, J.~A.~Chinellato and C.~Dobrigkeit,
  Astropart.\ Phys.\  {\bf 22} (2004) 159.

\bibitem{Dawson:2007di}
  B.~R.~Dawson  [Pierre Auger Collaboration], Proc. 30th ICRC (2007),
  arXiv:0706.1105 [astro-ph].

\end{thebibliography}
\end{document}